%% file: main.tex
\documentclass{article}

\usepackage[a4paper, margin=0.6in]{geometry}
\usepackage[utf8]{inputenc}
\usepackage[dvipsnames,table,xcdraw]{xcolor}
\usepackage[affil-it]{authblk}
\usepackage[misc]{ifsym}
\usepackage{multirow}

\usepackage{appendix,verbatim,listings,braket,enumitem, kantlipsum,graphicx,multicol,mathtools,csquotes,amsmath,mathtools,url,subfiles,qcircuit,amsthm,amssymb,fdsymbol,rotating,caption,hyperref,fontawesome,xcolor,arydshln,ulem,subcaption,caption, amsfonts}

\usepackage{pgfplots}
\usepackage{tikz}
\usetikzlibrary{shapes.geometric, arrows, positioning, backgrounds, shapes.arrows}
\usepackage{diagbox}
\usepackage{setspace}

\pgfplotsset{compat=1.18} 

\usepackage{xspace}
\newcounter{mynote}

\usepackage{sectsty}
\def\titlefont{\color{RoyalPurple}}
\sectionfont{\color{RoyalPurple}}
\subsectionfont{\color{RoyalPurple}}
\subsubsectionfont{\color{RoyalPurple}}

\let\OLDthebibliography\thebibliography
\renewcommand\thebibliography[1]{
  \small
  \OLDthebibliography{#1}
 \setlength{\itemsep}{0pt minus 0.3ex}
}

\title{\titlefont \textbf{\LARGE YAQQ: Yet Another Quantum Quantizer}\\ \Large Design Space Exploration of Quantum Gate Sets using Novelty Search}

\author[1,2,3, \Letter]{Aritra Sarkar}
\author[1,4,5,6 \Letter]{Akash Kundu}           
\author[2,3]{Matthew Steinberg}       
\author[2]{Sibasish Mishra}       
\author[2]{Sebastiaan Fauquenot}       
\author[1]{Tamal Acharya}           
\author[5]{Jaros\l{}aw~A.~Miszczak}   
\author[2,3]{Sebastian Feld}

\affil[1]{Quantum Intelligence research team}
\affil[2]{Quantum Machine Learning research group,
Quantum Computing division, QuTech, The Netherlands}
\affil[3]{Department of Quantum \& Computer Engineering, Delft University of Technology, The Netherlands}
\affil[4]{Joint Doctoral School, Silesian University of Technology, Gliwice, Poland}
\affil[5]{Institute of Theoretical and Applied Informatics, Polish Academy of Sciences, Gliwice, Poland}
\affil[6]{QTF Centre of Excellence, Department of Physics, University of Helsinki, Finland}

\affil[ \Letter ]{a.sarkar-3@tudelft.nl, akash.kundu@helsinki.fi}

\date{}

\begin{document}

\maketitle

\input{content}

\section*{Software availability} \label{software}

YAQQ is available on the Python Package Index (PyPI)~\cite{yaqqpip}.
The open-sourced code for YAQQ, configuration files, output data, and plotting codes for the experiments presented in this article are available at:

\href{https://github.com/Advanced-Research-Centre/YAQQ}{https://github.com/Advanced-Research-Centre/YAQQ}.

\section*{Acknowledgements}

AS thanks Pablo le Henaff for insightful discussions regarding Cartan decompositions and Zaid Al-Ars for discussions on using YAQQ as a compiler for NISQ processors. 
AS acknowledges funding from the Dutch Research Council (NWO) through the project ``QuTech Part III Application-based research" (project no. 601.QT.001 Part III-C—NISQ). 
AK and JM were partially supported by the Polish National Science Center (NCN) under the grant agreement 2019/33/B/ST6/02011. 
MS and SF are grateful for financial support from Intel. 

\section*{Author contributions}


Conceptualization, A.S.; 
Methodology, A.S., A.K. and T.A.; 
Software, A.S. and A.K.; 
Validation, A.K., A.S., S.M., M.S. and S.Fa.; 
Investigation, A.S., A.K., S.M. and M.S.; 
Writing – Original Draft Preparation, A.S., A.K. and M.S.; 
Writing – Review \& Editing, M.S., J.M., S.Fe., A.S., A.K., T.A.; 
Visualization, A.K., A.S., T.A, S.M. and S.Fa.; 
Supervision, S.Fe., J.M. and A.S.; 
Project Administration, A.S. and A.K.; 
Funding Acquisition, S.Fe. and J.M.

\newpage
\appendix
\input{appendix}   

\newpage
\bibliographystyle{unsrt}
\bibliography{ref.bib}

\end{document}

%% file: content.tex
\begin{abstract}
The standard model of quantum computation is based on quantum circuits, where the number and quality of the quantum gates composing the circuit influence the runtime and fidelity of the computation.
The fidelity of the decomposition of quantum algorithms, represented as unitary matrices, to bounded depth quantum circuits depends strongly on the set of gates available for the decomposition routine.
To investigate this dependence, we explore the design space of discrete quantum gate sets and present a software tool for comparative analysis of quantum processing units and control protocols based on their native gates. 
The evaluation is conditioned on a set of unitary transformations representing target use cases on the quantum processors.
The cost function considers three key factors: (i) the statistical distribution of the decomposed circuits' depth, (ii) the statistical distribution of process fidelities for the approximate decomposition, and (iii) the relative novelty of a gate set compared to other gate sets in terms of the aforementioned properties.
The developed software, called YAQQ (Yet Another Quantum Quantizer), enables the discovery of an optimized set of quantum gates through this tunable joint cost function. 
To identify these gate sets, we use the novelty search algorithm, circuit decomposition techniques (like Solovay-Kitaev, Cartan, and quantum Shannon decomposition), and stochastic optimization to implement YAQQ within the Qiskit quantum simulator environment.
YAQQ exploits reachability tradeoffs conceptually derived from quantum algorithmic information theory.
Our results demonstrate the pragmatic application of identifying gate sets that are advantageous to popularly used quantum gate sets in representing quantum algorithms.
Consequently, we demonstrate pragmatic use cases of YAQQ in comparing transversal logical gate sets in quantum error correction codes, designing optimal quantum instruction sets, and compiling to specific quantum processors. 
\end{abstract}

\section{Introduction} \label{section:intro}


Quantum computation is an emerging paradigm of computation that has captured the attention of both theoretical computer scientists and industrial computer engineers alike.
It is among the only known violations of the extended Church-Turing thesis~\cite{bernstein1993quantum}, allowing quantum algorithms to solve problems in specific complexity classes that are asymptotically intractable on all implementations of classical computation.
Many such quantum algorithms~\cite{qzoo} have been designed over the years, making quantum computers~(QC) a promising compute accelerator~\cite{sarkar2024automated} for these specific problems.  
A topical focus in QC is on engineering sufficiently robust quantum processors to demonstrate these computational benefits in practice, which has proved more difficult than anticipated~\cite{leymann2020bitter,ezratty2023we,waintal2024quantum}, leading to the idea of noisy intermediate-scale quantum~(NISQ) systems~\cite{preskill2018quantum} as a stepping stone to large-scale fault-tolerant quantum computation~(FTQC).
NISQ-era solutions focus on various tailored approaches~\cite{shi2020resource} to extract benefits from these limited computational devices.
These approaches include aspects at various layers of the QC stack~\cite{bertels2020quantum}, for e.g., pulse control~\cite{kairys2021efficient}, parametric quantum circuits and architecture search~\cite{patel2024curriculum}, error mitigation~\cite{botelho2022error} and correction~\cite{bhatnagar2023low}, circuit knitting on distributed architectures~\cite{chatterjee2022qurzon}, quantum circuit mapping and routing~\cite{steinberg2024resource}.
The research presented in this article applies to both NISQ and FTQC systems, albeit in different contexts.

Similar to classical computation, QC has many different models~\cite{wang2021comparative} that are polynomially equivalent: the quantum Turing machine~(QTM), adiabatic QC (AQC), one-way measurement-based QC~(MBQC) based on cluster states, and the most popular, gate-based QC~(GBQC) circuit model~\cite{deutsch1989quantum}.
In this work, we will focus on the GBQC model.
Inspired by the classical Boolean logic circuit model, GBQC comprises a set of initialized qubits representing a Hilbert space of equivalent dimension.
Quantum logic gates transform the states in that space, implementing the desired computation, and the qubits are eventually projectively measured to probabilistically extract the algorithm's output.


A crucial step in executing a quantum algorithm on a quantum processor unit~(QPU) is optimizing the quantum circuit, specified as a sequence of gates the quantum processor natively supports.
This optimization has a two-fold purpose.
Firstly, in the NISQ-era, the useful computational depth is bounded by the decoherence time of the qubits due to environmental coupling and the gate errors introduced by imperfect operations.
Longer unoptimized circuits lower the fidelity of the results for the algorithm due to these two factors. 
Secondly, the resource advantage of QC over classical models is typically quantified as asymptotic advantages in time complexity with respect to problem size.
Unoptimized circuits can considerably push the quantum-over-classical advantage further away in terms of problem size in both NISQ and FTQC scenarios.

Various approaches~\cite{wille2022basis} have been introduced to optimize quantum circuits, such as algebraic equivalences, ZX-calculus, variational compilation, unitary decomposition, binary decision diagrams, program synthesis, etc.
These optimization approaches often take into additional engineering and control constraints imposed by the underlying quantum system, such as qubit connectivity, native gate set, control frequency sharing, qubit manufacturing imperfections, etc.
To quantitatively validate the benefits of a holistic quantum application pipeline, quantum benchmarking suites and techniques have been proposed based on runtime, gate complexity, or fidelity, for e.g.,  QVolume~\cite{cross2019validating}, QPack~\cite{donkers2022qpack}, QScore~\cite{martiel2021benchmarking}, Quantum Resource Estimator~\cite{mykhailova2021testing} and MQT Bench~\cite{quetschlich2023mqtbench}, among others.

This research will focus on defining and quantifying quantum circuit dependencies related to the gate set decomposing the unitary operation.
To explore and numerically quantify this dependence, we present the Yet Another Quantum Quantizer~(YAQQ) \footnote{The YAQQ name is inspired by the YACC~\cite{johnson1975yacc}, a compiler-compiler, or a compiler-generator; similarly, YAQQ provides the decompositions for the generated gate set.} framework here.

This relation has been previously studied for theoretically inferring universal~\cite{barenco1995elementary} and optimal~\cite{parzanchevski2018super,kuperberg2023breaking} gate sets for QC or for pragmatically optimizing quantum compilation using native gate sets of specific QPU~\cite{lin2022let,davis2022gate,kalloor2024quantum} in a general-purpose QC formulation.
As presented in this article, the YAQQ approach deviates from previous approaches in three aspects.
Firstly, we consider a small and discrete native gate set, even though current QPU controls allow a richer set of operations.
Secondly, we consider a set of target unitaries as a benchmark that is used to conditionally evaluate the performance of various gate sets for the decomposition.
Thirdly, we do not specifically limit ourselves to evaluating currently available QPU gate sets and allow a hardware-agnostic discovery of novel gate sets that can, in principle, be engineered and characterized.
The rationale behind these design choices is elucidated in Section~\ref{section:design}.
Additionally, YAQQ does not access a quantum device at compile time and thus differs from methods~\cite{khatri2019quantum,jones2022robust} that access a quantum processor in an active learning feedback loop.


Besides the specific circuit employed to faithfully implement the quantum algorithm, the set of native gates supported by the underlying quantum processor, control electronics, and system software specifies an additional design degree of freedom that can be harnessed in both the NISQ and FTQC era.
This set directly affects the number of gates and the fidelity of the circuit composed of these gates.
To this end, the YAQQ framework allows the comparison of different QPUs (or quantum control firmware) based on the native gate sets they provide.
We benchmark the gate sets using the cost function, which takes into account three components:
(i) the circuit depth for the quantum transformation and their statistical distribution;
(ii) the process fidelity of the approximation of the quantum transformations and their statistical distribution and
(iii) the relative novelty of a gate set, with respect to a set of other gate sets, in the above two properties.
In this way, YAQQ conducts a design space exploration~(DSE) in this space of quantum gate sets and allows one to discover a set of quantum gates that optimize the scores based on a tunable joint cost function.
In conjunction with a novelty search~\cite{lehman2011abandoning}, YAQQ uses a generalized version of the Solovay-Kitaev theorem~\cite{dawson2005solovay}, random search, as well as SciPy-based optimization techniques to find these gate sets, within a Qiskit~\cite{Qiskit} quantum development environment.
As an example use case, we find a native gate set that would give a shallow circuit with high fidelity in decomposing unitaries.
In contrast, the typical fault-tolerant gate set (of Hadamard and T) gives a deeper circuit with relatively low fidelity when performing the same set of tasks. 
This can thereby be used to design an energy-efficient quantum instruction set architecture~(QISA).
We also demonstrate the application of YAQQ for a comparative analysis of transversal logical gates in quantum error correction~(QEC) codes.

In what follows, we present the main design choices for exploring quantum gate sets in Section~\ref{section:design}.
Section~\ref{section:dse_quantum_gate_sets} details the core algorithmic blocks used within YAQQ to perform the DSE.
Section~\ref{section:yaqq_sw} presents the YAQQ software framework.
In Section~\ref{section:results}, we discuss the results of fine-tuning the hyperparameters, discovering novel gate sets using a random dataset, and three exemplary applications of YAQQ.
Section~\ref{section:conclusion} concludes the article.

\section{Theoretical rationale behind design choices} \label{section:design}

In this section, we present the three design choices that set YAQQ apart from related work and the rationale behind these choices.

\subsection{Abstraction level of quantum unitary control and synthesis}


At an abstract level, GBQC can be expressed as a single $n\times n$ unitary gate $U$ acting on a known $n$-qubit initial state (typically, $\ket{0}^{\otimes n}$ product state) and measured out in the computational $Z$-basis.
This $U$ can be optimized in three levels of abstraction: (i) control signals for a single unitary, (ii) sequence of local operations from a parametric family of operations, and (iii) sequence of local operations from a finite set of operations.
These levels are explained herein.

\subsubsection{Directly optimizing a large unitary}

The time required to implement a unitary directly relates to the energy expenditure for realizing the corresponding quantum process, and thus, it is sensible to optimize for time.
Quantum speed limits~(QSL) provide a lower bound on the time required to transport an initial quantum state to a final one through a physical process, e.g., subjected to external control fields.
QSLs can be defined in many different ways~\cite{pires2016generalized}.
The Mandelstam-Tamm bound~\cite{tamm1991uncertainty} $T_{MT} = \dfrac{\hbar \Theta_{0T}}{\Delta H}$ provides an operational interpretation of the energy-time uncertainty relation; while, the Margolus-Levitin bound~\cite{margolus1998maximum} $T_{ML} = \dfrac{\hbar \Theta_{0T}}{\braket{H}}$ depends on the expectation value $\braket{H}$ of the generator of time evolution, rather than its variance $\Delta H$. 
The unified bound, $T \ge T_{QSL} = \max\{T_{MT}, T_{ML}\}$ is tight~\cite{levitin2009fundamental}.
These bounds are expressed using the Hilbert space angle $\Theta_{0T} \equiv \arccos(\braket{\psi_0|\psi_i})$ between the initial state $\ket{\psi_0}$ and the final state $\ket{\psi_i}$.
Note that QSL is defined for states and not unitaries, and thus, the same unitary operator acting of different states, e.g., $X\ket{0} = \ket{1}$ vs. $X\ket{+} = \ket{+}$ will have different bounds.

QSL can be understood from a purely geometric perspective~\cite{anandan1990geometry,dowling2008geometry} as a consequence of the quantum geometric tensor's~(QGT) properties (also called the Fubini-Study metric)~\cite{cheng2010quantum}.
The real part of QGT imposes a Riemannian metric on a manifold that measures the quantum distance, while the imaginary part is the Berry curvature.
The Bures angle gives the geodesic length, and the path length of any unitary dynamics for a time-dependent Hamiltonian is given by $\int_0^{T} dt \Delta H(t) / \hbar$.
Geometric QSLs~\cite{pires2016generalized} are insensitive to the speed at which a path is traversed; they only rely on the time-averaged speed.
Action QSL~\cite{o2021action} generalizes this by showing that the optima depends on the path taken and the speed at which that path is traversed. 
The best way to traverse a path is tackled with quantum optimal control theory, which can be used to optimize the energy cost of quantum computation~\cite{aifer2022quantum}.

An alternate research avenue focuses on the speed at which information propagates in quantum spin systems. The effective light cone determines the Lieb-Robinson bound~\cite{lieb1972finite}, beyond which information propagation exponentially decays with increasing distance.
QSLs are primarily concerned with determining the shortest time required for state changes but may not provide accurate predictions for many-body systems~\cite{bukov2019geometric}, while the Lieb-Robinson bound offers valuable insights into many-body systems but does not furnish information regarding the speed of transitions inside the light cone.
These two bounds can be unified~\cite{van2023topological} by harnessing the capabilities of optimal transport theory.


While QSL and QGT provide the optimal set of controls for implementing $U$, admittedly, in a QPU, the number of control Hamiltonians that can be driven in tandem is restricted by various factors like control bandwidth and crosstalk induced by frequency couplings.
The minimal conditions of controllability of a closed quantum system for a set of control Hamiltonians are given by the Lie rank test~\cite{dirr2008lie}.
While general quantum control is computationally hard~\cite{bondar2020uncomputability}, the field of optimal quantum control~(OQC) provides various strategies and heuristics to design the control pulses to implement a target $U$.
These strategies not only restrict reaching the QSL to accessible geodesics~\cite{bukov2019geometric} but, more importantly, limit the maximum size of $U$ that can be implemented without composing out of smaller logical blocks.
Since this is the only scalable strategy for large-scale QC within YAQQ, we do not consider optimizing at the pulse level and focus on decomposing $U$ into smaller building blocks composing a rich gate set, although one could, in principle, examine this in the future.
Each gate within a gate set is assumed to require a unit time step.
Some hardware-aware compilers take into account the gate execution time of the target hardware.
To generalize this runtime-aware circuit decomposition, in Section~\ref{app:pulse} we show how the results from YAQQ can be further optimized at the pulse level, which can then be incorporated in a closed-loop (time or energy) resource estimate for the elements within a gate set.

\subsubsection{Optimizing a unitary using a family of parametric gates}

A set of quantum logic gates is considered computationally universal if any quantum computation can be efficiently expressed using those gates.
The $U$ can be decomposed into $k$-local gates, i.e., using gates that act on at most $k$ qubits.
Note that this is irrespective of the physical locality on the QPU.
The routing process ensures the physical locality of the $k$ qubits involved in these $k$-local gates and introduces a worst-case constant factor overhead~\cite{steinberg2024resource} to the runtime dependent on the QPU size. 
Finding the minimum value of $k$ for universal QC has been of both theoretical and engineering interest.
For example, the Deutsch gate $D(\theta)$ of $k=3$ is a single parametric gate that is universal for QC.
Universal QC can be minimally achieved~\cite{lloyd1995almost} with $k=2$, such as the $2$-qubit gate CX-gate and single-qubit arbitrary angle rotation gate along any $2$ mutually orthogonal axes.
It has been proven that an exact decomposition of an arbitrary $n$-qubit gate requires at least $\frac{1}{4}(4n-3n-1)$ CX gates~\cite{krol2024highly,shende2004minimal}.
This worst-case exponential cost for $k$-local universal QC implies that a small subset of unitaries will be practical for expressing and executing quantum algorithms.
Note that similar worst-case exponential circuit complexity results also hold for classical $2$-local Boolean circuits~\cite{shannon1949synthesis}.
The corresponding algorithm to decompose $U$ using is aptly named Quantum Shannon Decomposition~(QSD)~\cite{shende2005synthesis}.
QSD is an important first-order decomposition that can easily be recursively generalized to $n$-qubit and forms an important primitive within this work.

\subsubsection{Decomposing a unitary using a finite set of discrete gates} \label{sec2p1p3}


Once the $2$-local quantum gates are routed to be physically nearest neighbors on the QPU's qubit connectivity (typically by additional SWAP gates \cite{steinberg2024resource}), current QPU models support universal QC via a $2$-qubit entangling gate (e.g., CX or CZ) along with arbitrary rotation along X, Y, Z axes.
Despite this capability, we consider a finite set of discrete gates for the decomposition in this work.
We justify this choice threefold. 

Firstly, quantum characterization is exponentially costly in terms of resources.
Full characterization, for example, via gate set tomography~\cite{yu2024transformer}, is typically done for a small set of rotation angles (e.g., $90^\circ$ and $45^\circ$ for each of the $3$ axes).
The characterization is, in turn, used to tune the control electronics and the compilation process; thus, only the characterized subset of gates can be reliably used for the QPU.

Secondly, FTQC necessitates employing QEC, which encodes the quantum information of a single qubit in a set of physical qubits called a logical qubit.
This encoding then performs a universal set of operations at the logical level by local operations on the physical level.
Some operations that can be easily translated to this local form while maintaining the fault tolerance of the QEC are termed transversal gates.
Transversality is typically proven for specific discrete gates using a specific QEC, instead of a family of parametric gates.
Moreover, due to the Eastin-Knill theorem~\cite{eastin2009restrictions}, it is known that transversal gate sets cannot be universal and require additional resources like magic state distillation.
Thus, FTQC will be composed of a small set of transversal gates for the chosen QEC and additional resource states.
Thereby, every $U$ needs to be decomposed into this set of discrete operations with maximum fidelity before introducing additional resource states.

Thirdly, this choice can be justified even in a NISQ control setting, where we assume that the QC is operated for arbitrary parametric gates (irrespective of the characterization data being available or used in the control) without a QEC code.
NISQ-era algorithms focus on variation quantum circuits operated in a hybrid loop of low-depth parameterized quantum ansatz and a classical parameter optimizer.
The success of these algorithms requires a continuous parametric space~\cite{sayginel2023fault}, which gets severely limited by quantum stack layers connecting the QPU to the classical optimizer.
The precision of these parameters is typically discretized by the quantum programming language's datatype encoding the angle, the microarchitecture's quantum instruction bandwidth, and the digital-to-analog converter's resolution for microwave pulse control.
While the set of discrete controllable gates is considerably large in this setting, as the size of the Hilbert space grows with larger quantum systems, the reachable volume of the Hilbert space within bounded errors will still exponentially reduce, considering the $k$-locality of these gates from QSD.

The usual convention in formal languages and automata theory is that the set of alphabets is countable and finite.
This matches well with our classical intuition and natural languages.
In universal computation models, like the Turing machine, this restriction is associated with a finite-size program, represented as a finite-state machine, which can effectively use only a finite alphabet.
Research in automata theory for languages with countable infinite alphabets (e.g., over integers $\mathbb{Z}$) is sparse and relatively recent~\cite{manuel2012automata}.
However, it is known that models of computation over uncountable infinite sets supported by unconventional physical settings (e.g., over real numbers $\mathbb{R}$, or using closed time-like curves) are more powerful than Turing machines.
Such models allow hypercomputation~\cite{aaronson2005guest} (e.g., calculating the halting probability or exact solutions to NP-Hard problems) as the size of the sets is related to their computational power via the arithmetic hierarchy.
As an initial intuition, the Hilbert space where quantum computation happens is a continuous complex-valued vector space.
For finite-dimensional quantum systems, e.g., qudits (qubits when $d=2$), this would match the set $\mathbb{C}^d$, and should be at least as powerful as real computing.
However, we know that quantum computers, formalized as quantum Turing machines~(QTM), lie in the same arithmetic hierarchy as classical universal Turing machines~(UTM).
Thus, every function that is computable with QTM is also computable with UTM, albeit sometimes requiring exponential time/memory resources, e.g., for complexity classes like EQP instead of P (or BQP instead of BPP).
To internalize this, it is crucial to understand that, in the definition of QTM, while the inputs can be a superposition of basis states (i.e., from $\mathbb{C}^d$), the program is classical and composed of a finite language~\cite{sarkar2020quantum}.
The generalization of hypercomputation using quantum mechanics indeed exists in the form of quantum field computation~\cite{manoharan2001unity}, a largely theoretical research direction at present.
Thus, while quantum computation is orchestrated in the uncountable continuous space, quantum programming and control are fundamentally limited by human interfacing languages expressed using a finite discrete set.
Whether this nullifies the quantum computational resource advantage with respect to classical algorithms is debatable~\cite{pfister2013information}.


In QC, the set of universal gate sets is uncountably infinite. Hence, it is a richer counterpart to the ubiquity of universality~\cite{wolfram2002new} in classical computation.
We refer to this as the ubiquity of quantum universality~(UQU).
The Solovay-Kitaev decomposition~(SKD) allows one to decompose an arbitrary $1$-qubit $U$ using the discrete gate set of \{H, T\}.
In its general form, the corresponding Solovay-Kitaev theorem~(SKT) states that for a finite set of elements $\mathcal{G}$ in the special unitary group $SU(2)$ containing its inverses and generating a dense group, for some precision error $\epsilon > 0$; for any $U \in SU(2)$ there is a sequence $S$ of gates in $\mathcal{G}$ of length $O(\log^{\log(5)/\log(3/2)}(1/\epsilon))$ (Equation~8 in \cite{dawson2005solovay}) such that the operator norm error is bounded $||S-U|| \le \epsilon$.
Moreover, the decomposition can be computed in $O(\log^{\log(3)/\log(3/2)}(1/\epsilon))$ time (Equation~9 in \cite{dawson2005solovay}).
$\mathcal{G}$ denotes the group generators (a finite subset of $SU(d)$)~\cite{ozols2009solovay}.
The corresponding generators of the Lie algebra, $\mathfrak{su}(d)$, are the Hermitian matrices denoting the Hamiltonians for these generators (e.g., the Pauli matrices for $\mathfrak{su}(2)$ and the Gell-Mann matrices for $\mathfrak{su}(3)$).

It must be noted that both the decomposed circuit length and the time to decompose scales exponentially with the number of qubits (Section~5.1 in \cite{dawson2005solovay}).  
While SKT will always find a good approximation for any value of $\epsilon$, the search covers only a very sparse region of the entire space of possible approximation sequences. 
Consequently, the approximation's output is almost always far longer than it needs to be~\cite{trung2012optimising}.
Thus, as an alternative, we employ random decomposition~(RD), which creates sequences of random length with a random sample of gates from the gate set and chooses the sequence with the highest fidelity among the set number of trials.
The advantage of RD is that it can be easily extended to higher Hilbert spaces, though the number of random trials needs to be adjusted accordingly.

In this work, we use a combination of QSD from decomposing $n$-qubit unitaries to $2$-local gates, Cartan decomposition for converting between CX and available $2$-local gates in the target gate set, and SKD for the final $1$-qubit decomposition. RD is made available as an alternative stochastic strategy for all these steps.
The performance for these strategies is presented in Section~\ref{section:results}.

\subsection{Dataset of unitaries for benchmarking} \label{section:design_data}


In contrast to circuit compilation, YAQQ is not concerned with optimizing one specific unitary.
We require the gate set to work reasonably well for all/most algorithms in terms of both expressibility and fidelity.
Thus, the comparative evaluation of the gate sets is conditioned on a set of unitaries.
This dataset can be a general Haar random set of unitaries or a curated set of unitaries for specific quantum algorithms or states that need to be prepared.
The dataset strongly influences the comparative scores of the gate sets in YAQQ for two reasons: interpretability and reachability.

Typically, gate sets used in quantum computing are based on physical interpretations of quantum phenomena, e.g., bit-flip, phase-flip, and controlled-NOT.
Thus, the gate sets also serve as composable blocks for intuiting quantum algorithm design.
A large set of gates loses explainability~\cite{arrieta2020explainable} in the sub-symbolic black-box paradigms like neural networks and variational algorithms, thus the requirement of a finite and discrete gate set is motivated also from the interpretability perspective.
While YAQQ does not natively harness these compositional blocks due to our focus on the QPU native gate set, YAQQ can be used to discover algorithmic building blocks.
However, YAQQ's gate set search overlaps the functionalities of quantum circuit element discovery (as discussed in \cite{trenkwalder2022automated,sarra2023discovering}) and quantum program synthesis~\cite{sarkar2022automated}.   
To this end, in Section~\ref{app:qisa}, we present a framework to employ YAQQ for interpretable quantum gate set discovery and the development of efficient QISA and quantum intermediate representations~(QIR).

The DSE of YAQQ over quantum gate sets is based on reachability.
While universality is easy to achieve in quantum computing~\cite{lloyd1995almost}, the core thesis of YAQQ is to compare gate sets that are universal by more pragmatic bounded resources in reachability (instead of expressibility). 
To clarify, expressivity refers to the extent to which the Hilbert space can be encoded using an unbounded number of gates.
All universal gate sets share the same asymptotic expressibility unto polynomial resource overheads.
Reachability refers to a bounded form of expressibility~\cite{bach2023visualizing}, such that the length of the sequence of gates is shorter than the specified bound, which typically corresponds to the decoherence time of the processor.
Thus, expressibility is dictated by computability, while reachability is dictated by complexity.
For example, let the gate set be $\{G_A,G_B\}$.
These gates' individual (informational and engineering) complexity is assumed to be comparable.
Thus, using this gate set, the states reachable is $1$ time step is $\{G_A,G_B\}$, in $2$ time step is $\{G_A.G_A,G_A.G_B,G_B.G_A,G_B.G_B\}$, and so forth.
Another gate set, $\{G_C,G_D\}$, would likely reach very different states in a bounded number of time steps.
Besides the dependence on the circuit depth, the size of the Hilbert space grows much faster than the number of reachable points (and their neighboring bounded-error approximation space).
Thus, if the circuit depth is bounded, gate sets will significantly diverge in their reachability space for larger unitaries and thus can be harnessed by YAQQ to propose an optimal gate set.
This dependence is reminiscent of no-free-lunch-theorems~(NFLT)~\cite{schaffer1994conservation,wolpert1997no} in optimization and machine learning.

Conversely, each unitary has a preferred gate set (e.g., the unitary itself, which would reach it in a single step).
Additionally, given two universal gate sets, the number of gates required to perform unitary decomposition within a specified approximation error bound would differ only in logarithmic factor due to SKT, as discussed before.
However, since the circuit depth for bounded error scales exponentially with larger Hilbert spaces (also due to SKT), larger unitaries' data set strongly influences the gate sets' resource requirements to achieve comparable approximation error bounds.

In this work, we consider various types of data sets to evaluate the gate sets in YAQQ.
It is, however, easy to customize YAQQ with a use-case-specific data set.
These options are further discussed in Section~\ref{sec:data_set}.

\subsection{Hardware-agnostic novelty search of gate sets}


YAQQ discovers a set of quantum gates that optimize the tunable joint cost function.
This joint function comprises the approximation fidelity, the circuit depth, and the novelty score.
As discussed, this optimization is conditioned on the set of target unitaries.
The novelty component of the cost function is discussed herein.

Novelty search (NS)~\cite{lehman2010efficiently,lehman2011abandoning} is an evolutionary algorithm that prioritizes discovering diverse and novel solutions over-optimizing for a specific objective.
As opposed to extrinsic rewards, it is a technique to score intrinsic motivation~\cite{baldassarre2013intrinsically}.
NS encourages exploration by rewarding solutions that are dissimilar from those already discovered, regardless of their performance, thus avoiding local optima and promoting the discovery of innovative solutions that traditional fitness-based algorithms might overlook.
This approach has applications in various domains, including robotics, game design, and optimization problems, where creativity and diversity are valued alongside performance.
Novelty search is particularly useful in complex problem domains where the search space is large and traditional optimization techniques may struggle to find the global optima.
While novelty is beneficial for these reasons, we cannot deny the eventual goal of efficiency of the quantum gate set for quantum compilation and control.
Therefore, we balance these two factors by including a novelty score to the cost function~\cite{mouret2011novelty}, which can tune the performance-novelty tradeoff.


There is a subtle difference between using a finite set of operators for QC and classical Boolean circuit optimization.
In a classical setting, all functions can be perfectly represented by a sequence of gates from the universal gate set $G$.
In contrast, the quantum setting aims to approximate all possible unitary operations with a sequence of gates from $G$ with a bound of the approximation quality.
The quantum case can be understood by drawing parallels to representing all real numbers using digits of a specific numeral base.
Thus, an obvious trade-off exists in taming this countably infinite space with a finite number of building blocks.
This tradeoff is formalized in quantum algorithmic information theory~(QAIT), specifically via quantum Kolmogorov complexity~(QKC)~\cite{vitanyi2001quantum,berthiaume2001quantum}, which measures the information required to describe a quantum state. 
By simple counting arguments from QAIT~\cite{mora2007quantum}, it can be shown that for any pure quantum states described by a $b$ classical bits using a specific encoding scheme, at least another pure state of the same dimension exists that is incompressible using the same scheme.
This principle can be generalized to quantum unitaries~\cite{kaltchenko2021kolmogorov} and is the underlying reason why the novelty search on gate sets can be expected to discover gate sets that are at least similar in encoding efficiency conditioned on the set of target unitaries.
Further details on the construction of the joint cost function are explained in Section~\ref{section:dse_quantum_gate_sets}.



\section{Design-space exploration of quantum gate sets} \label{section:dse_quantum_gate_sets}




YAQQ is a tool for automated quantum gate set design.
The target is to develop specifications for a quantum processor that is yet to be manufactured (or the operating procedures of a tunable quantum computer). 
However, it can also be used for a comparative study of existing specifications or an in-depth evaluation of a single specification.
We are interested in the space of building blocks that allows efficient gate arrangements for building practical quantum applications and the circuit structure based on those blocks.
For an effective DSE, five aspects of the problem formulation are crucial.
These aspects, as detailed in the rest of this section, are: (i) data set for comparing the gate set, (ii) cost function for evaluating the novel gate set, (iii) circuit decomposition technique for using the gate set in compiling unitaries, (iv) search technique for finding the novel gate set, and (v) novel gate set ansatz for pruning the search space.


\subsection{Data set}\label{sec:data_set}

\begin{figure}[!ht]
     \centering
     \begin{subfigure}[b]{0.22\textwidth}
         \centering
         \includegraphics[width=\textwidth]{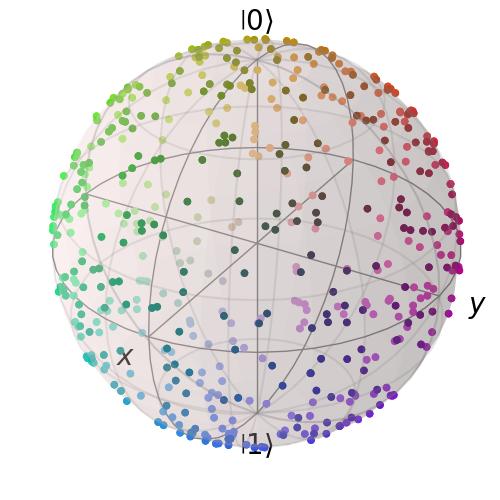}
         \caption{Haar random states}
         \label{fig:ds_randS}
     \end{subfigure}
     \begin{subfigure}[b]{0.22\textwidth}
         \centering
         \includegraphics[width=\textwidth]{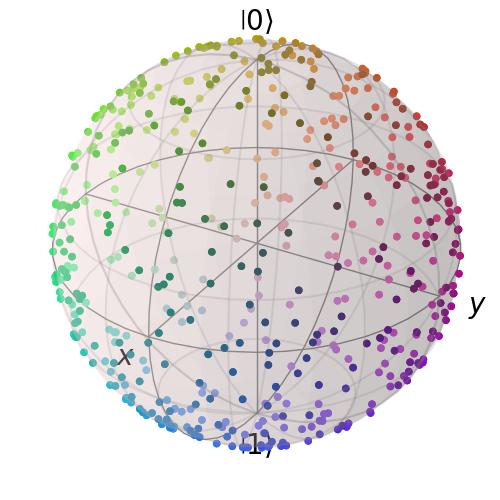}
         \caption{Haar random unitaries}
         \label{fig:ds_randU}
     \end{subfigure}
     \begin{subfigure}[b]{0.22\textwidth}
         \centering
         \includegraphics[width=\textwidth]{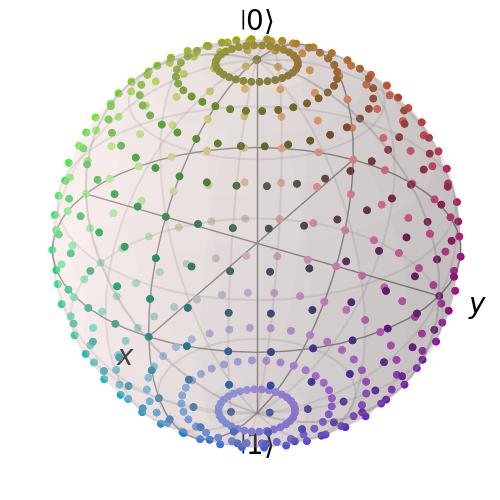}
         \caption{Equispaced angles}
         \label{fig:ds_fibo}
     \end{subfigure}
     \begin{subfigure}[b]{0.22\textwidth}
         \centering
         \includegraphics[width=\textwidth]{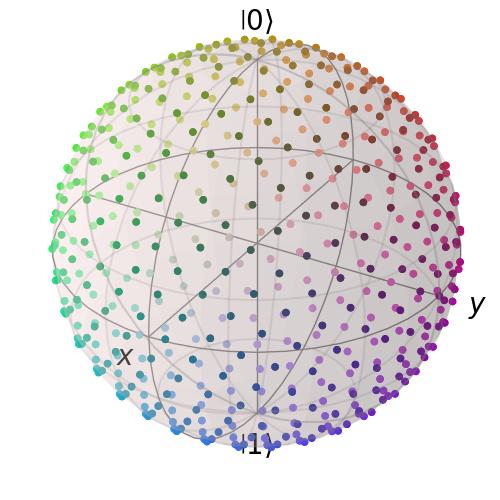}
         \caption{Equispaced points}
         \label{fig:ds_equi}
     \end{subfigure}
        \caption{1-qubit Data Sets of size 512}
        \label{fig:ds1q}
\end{figure}

\begin{figure}[thb]
     \centering
     \begin{subfigure}[b]{0.48\textwidth}
         \centering
         \includegraphics[clip, trim=2cm 1.5cm 2.5cm 4.0cm,width=\textwidth]{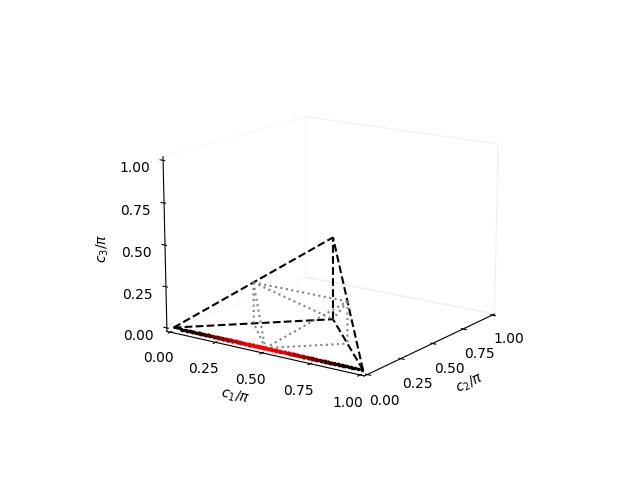}
         \caption{Haar random states}
         \label{fig:ds_equiWeyl}
     \end{subfigure}
     \begin{subfigure}[b]{0.48\textwidth}
         \centering
         \includegraphics[clip, trim=2cm 1.5cm 2.5cm 4.0cm,width=\textwidth]{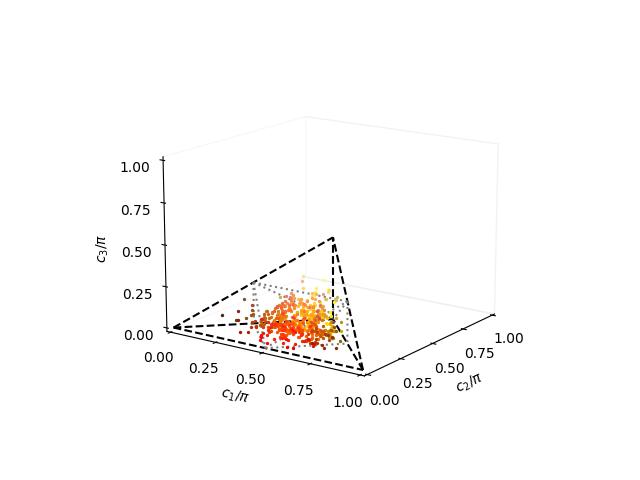}
         \caption{Haar random unitaries}
         \label{fig:ds_randWeyl}
     \end{subfigure}\\
     \begin{subfigure}[b]{0.48\textwidth}
         \centering
         \includegraphics[clip, trim=2cm 1.5cm 2.5cm 4.0cm,width=\textwidth]{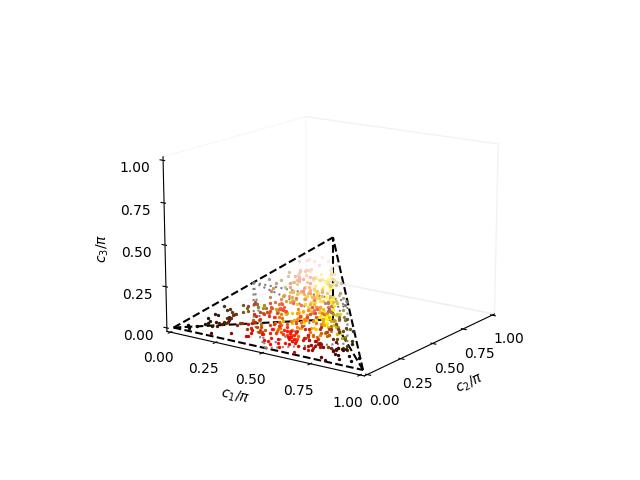}
         \caption{Random Weyl chamber parameters}
         \label{fig:ds_haarWeyl}
     \end{subfigure}
     \begin{subfigure}[b]{0.48\textwidth}
         \centering
         \includegraphics[clip, trim=2cm 1.5cm 2.5cm 4.0cm,width=\textwidth]{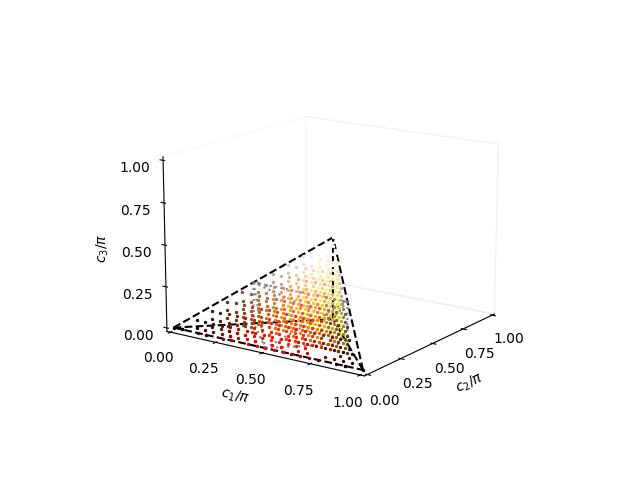}
         \caption{Equispaced non-local}
         \label{fig:ds_equiWeylNL}
     \end{subfigure}
     
        \caption{2-qubit Data Sets of size 508. The axes corresponds to the $t_x, t_y, t_z$ parameters in the 2-qubit canonical gate as defined in Equation~\ref{eq:nl2} in Section~\ref{s2p5}.}
        \label{fig:ds2q}
\end{figure}

The data set, as introduced in Section~\ref{section:design_data}, refers to the set of unitaries used for the comparative analysis of the gate sets.
Here, we present the data set options available within YAQQ.
Besides these options, as we demonstrate in Section~\ref{sec:apps}, a use-case-specific data set can also be easily supplied for the DSE.
Datasets for $1$- and $2$-qubit can be visualized on the Bloch sphere and Weyl chamber, respectively.
Typical experiments in quantum information are performed by sampling unitaries from the Haar measure.
YAQQ uses the default data set as Haar random unitaries of a specified sample size for an $n$-qubit Hilbert space dimension.
State preparation unitaries for Haar random states can also be selected.
The Haar random state and Haar random unitary should give similar results, as in principle, a Haar random unitary on any state (including the $\ket{0}$ state) is a Haar random state\cite{maziero2015random}. 
This relation between states and unitaries differs from discrete classical~\cite{vitanyi2001quantum} or discrete quantum circuits~\cite{bach2023visualizing}, where the universal distribution is generated for random sampling of gates or programs.
For $1$-qubit, we provide additional options: (a) equispaced points on the Bloch sphere generated by a golden ratio rotation on the spherical coordinate, (b) uniformly spaced parameters for IBM U3 gate (refer \cite{qiskitU3}, discussed later in Equation~\ref{eq:u3}), and (c) stabilizers and magic states for evaluating transversal logical gate sets in quantum error-correction codes.
For 2-qubit, we provide additional options: (a) equispaced non-local gates on the Weyl chamber and (b) uniformly spaced parameters for Weyl chamber coordinates.
Figures~\ref{fig:ds1q} and \ref{fig:ds2q} visualize the 1-qubit and 2-qubit dataset options.
All datasets are stored as unitaries. 
The data set, defined as a set of unitary matrices, can be converted to a set of superoperators, e.g., the Choi matrix~\cite{jiang2013channel}. 
This conversion permits the future applicability of YAQQ to noisy quantum compilation. 


\subsection{Cost function}\label{sec:cost_fn}

The cost function for the DSE compares two quantum gate sets conditioned on the data set.
We use a multi-modal cost function defined as a weighted sum of $5$ metrics.
\begin{equation}
    c~(gs_2|gs_1,ds) = 
    \underbrace{w_{apf}.c_{apf} + w_{npf}.c_{npf} + w_{acd}.c_{acd} + w_{ncd}.c_{ncd}}_{\text{quantum information}} \;\;\;\;\;\;+\underbrace{ w_{agf}.c_{agf}}_{\text{quantum engineering}}
\label{eq:cost_function}
\end{equation}

The first metric $c_{apf}$ is the improvement of the average process fidelity from the first gate set to the second as a metric for rewarding `better' decompositions.
The process fidelity~(PF) between two decomposed quantum circuits is calculated in the superoperator form as $\text{PF}(qc_1,qc_2) = \text{Tr}[\chi_{qc_2}^\dagger,\chi_{qc_1}] / d^2$, where $\chi$ is a superoperator representation of the unitary corresponding to the quantum circuit, and $d$ is the system dimension.
The second metric, $c_{npf}$, is the novelty score in the process fidelity.
The novelty of a gate set corresponds to a reciprocal trend of the process fidelity compared to the process fidelity obtained with another gate set, in the task of decomposing random unitaries.
To measure this novelty score, for each data point, the process fidelity by about $0.5$ to capture the trend (i.e., which states have lower fidelities in gate set 1).
Thereafter, the distance between this trend is measured and normalized by the average fidelity of the gate sets.
In the remaining two metrics, we improve the average depth and the novelty in depth of the decomposition.
We do not consider extra ancilla qubits for decompositions.
The circuit depth is measured as a unified cost for each quantum gate irrespective of the number of qubits, and not the critical path cost, thus characterizes an upper bound of the runtime.
The third metric $c_{acd}$ is the improvement of the average circuit depth from the first gate set to the second as a metric for rewarding `frugal' decompositions.
The fourth metric, $c_{ncd}$, measures the novelty in terms of circuit depth and gives a higher preference to gate sets that perform decompositions with a `novel' trend in circuit depth, thus giving shorter circuits that are longer in gate set 1.

The two metrics corresponding to novelty, $c_{npf}$ and $c_{ncd}$, together allow for concept discovery on the data set.
The overall cost function in the first four metrics allows tuning a tradeoff between novelty and objective \cite{mouret2011novelty}.
This is the first time novelty search has been explicitly used in quantum computing (or automata modeling), though other forms of intrinsic motivation~\cite{baldassarre2013intrinsically} have been explored in \cite{trenkwalder2022automated, sarra2023discovering,sarkar2022qksa}.

While the first four metrics are quantum information-theoretic in nature, the reason to consider a particular gate set is often dictated by the ease of fabricating and controlling the physical quantum processor in implementing the operation.
Thus, the fifth metric $c_{agf}$ can bias the cost function by considering the average gate fabrication or control difficulty.
This value can be easily embedded within YAQQ based on the engineering perspective of a specific QPU.

\subsection{Decomposition techniques}\label{sec:decomp_tech}

The decomposition of the target unitary data set via a candidate gate set in the DSE can be performed using different techniques that the user can specify.
There are four decomposition techniques, each applicable for different sizes of unitaries: (i) Solovay-Kitaev decomposition~(SKD) for $1$-qubit unitaries to a discrete and finite set of $1$-qubit gates, (ii) Quantum Shannon decomposition~(QSD) for $n$-qubit unitaries to CX and additional $1$-qubit rotation gates along RY and RZ, (iii) Cartan decomposition~(KAK) for replacing $2$-qubit gates with some other $2$-qubit gate (at max. $4$) and additional $1$-qubit rotation gates, and (iv) Random decomposition~(RD) for $n$-qubit unitaries to a discrete and finite set of $k$-qubit gate, such that $k \le n$.
These decomposition techniques can be used in combination, for example, using QSD for $n$-qubit unitaries, and then using KAK to decompose CX to available $2$-qubit gates in the gate set, and finally the $1$-qubit unitaries from QSD and KAK decomposed with available $1$-qubit gates in the gate set using SKD or RD.
In this section we briefly review the four decomposition techniques.


\subsubsection{Solovay-Kitaev decomposition}

The SKD algorithm~\cite{dawson2005solovay} applies the Solovay-Kitaev theorem to decompose quantum operators into a sequence of quantum gates. 
As explained in Section~\ref{sec2p1p3}, the algorithm finds approximate sequences such that the length of the decomposed circuit and the runtime of the algorithm scales as a logarithmic factor with the inverse precision $\dfrac{1}{\epsilon}$.
SKD includes a preprocessing step that generates a search space of composite sequences up to a length (or depth) $d$ of gates that belong to a finite discrete basis. 
This set of composite sequences is referred to as the SK-basis.
The algorithm functions in a recursive fashion, and the degree of recursion is denoted by $n$.
The algorithm returns a sequence that approximates the unitary operator $U$ to an error $\epsilon_n$ during the process. 
The algorithm is designed to obtain an improved approximation accuracy $\epsilon_r < \epsilon_{r-1}$, and eventually, the base case returns the \texttt{best\_approximation} to a matrix $U$, bounded by $\epsilon_0$. 
The detailed steps of each recursion step are explained in Appendix~\ref{a1}. 
The approximation accuracy tends to reach 0 as the recursion depth $n$ increases indefinitely. 
In YAQQ, the qiskit implementation of SKD has been modified to allow arbitrary single-qubit unitary gates (along with its inverse/dagger) to form the gate set for the SK-basis.
Such a modification allows us to employ SKD in the DSE of candidate gates in the gate sets.

An extension of the algorithm for single-qubit systems to the general case of $m$-qubit systems is possible in theory~\cite{dawson2005solovay} with the modification in only one step of the algorithm -- the balanced commutator group decomposition to an approximate version of itself. 
While this change appears reasonably straightforward, it has a significant non-trivial implementation.
The depth of the SK basis also scales exponentially with the dimension of the system. 
Therefore, generalizing SKD to higher dimensions doesn't function well, necessitating Quantum Shannon Decomposition, an exact decomposition technique covered subsequently.

\subsubsection{Quantum Shannon decomposition}

QSD was proposed~\cite{shende2005synthesis,krol2022efficient} as a technique for expressing any $n$-qubit quantum operator as an exact decomposition into single-qubit rotations and 2-qubit controlled gates.
The algorithm follows a divide-and-conquer strategy in a recursive fashion and breaks down the $n$-qubit unitary matrix $U$ into smaller sub-matrices. 
QSD starts with cosine-sine decomposition~(CSD), a well-known technique in linear algebra that divides the target matrix $U$ into smaller blocks. 
It then recursively performs CSD~\cite{tucci1999rudimentary} and other decomposition techniques, such as eigenvalue decomposition and Euler decomposition, to eventually express the original complex operator as a sequence of single-qubit gates and CNOTs.
This synthesis technique is a quantum version of the classical Shannon decomposition of boolean functions. 
The QSD steps are explained in more detail in Appendix~\ref{a1}.
The single qubit gates and CNOT from the QSD can be further decomposed via SKT or random decomposition to the available gate set.

\subsubsection{Cartan decomposition} 

The QSD decomposes $n$-qubit unitaries to CNOT and single-qubit rotations.
The single-qubit rotations can further be decomposed to the target gate set via SKD.
However, the target gate set might not include CNOT.
This constraint necessitates using Cartan decomposition (KAK)~\cite{tucci2005introduction}.
The acronym KAK~\cite{khaneja2001cartan} refers to the use of a group $\mathcal{G} = \exp(g)$ with a subgroup $\mathcal{K} = \exp(k)$ and a Cartan subalgebra $a$, where $g = k \oplus k^\perp$ and $a \subset k^\perp$.
Any $G \in \mathcal{G}$ can be expressed as $G = K_1 A K_2$, where $K_1, K_2 \in \mathcal{K}$, and $A \in \exp(a)$. 
This special case of Cartan Decomposition allows one to factor a general $2$-qubit operation (i.e., an element of $U(4)$) into local operations applied before and after a $3$-parameter, non-local operation. 
A general $2$-qubit gate corresponds to a $4\times 4$ unitary matrix with $16$ free parameters ($15$ considering an irrelevant phase).
Based on KAK decomposition, any $2$-qubit gate can be expressed as a canonical gate, plus $4$ local $1$-qubit gates, thus tuning the $15 = 3 + 4*3$ parameters.
The canonical gate is defined in Section~\ref{s2p5}.
This decomposition based on the canonical gate is known as the magic- or Kraus-Cirac- decomposition~\cite{kraus2001optimal} and is explained in more detail in Appendix~\ref{a1}.
Note that if the gate set provided does not have a perfect two-qubit entangler, the qiskit implementation of approximate KAK decomposition would still return a circuit with the highest fidelity, though it could be rather low.

\subsubsection{Random decomposition} 

RD is a flexible alternative for decomposing general unitary matrices using an arbitrary gate set.
In RD, a set of sequences of random length with a random sample of gates from the gate set is evaluated.
The sequence with the highest fidelity among the set number of trials is returned as the decomposed circuit.
Though it sounds naive, in practice, this performs better than SKT for single qubit decomposition on most target unitaries. 
While the advantage is that it can be easily extended to higher Hilbert spaces, the number of random trials needs to be adjusted accordingly and will most likely grow prohibitively for large unitaries.

\subsection{Search technique}\label{sec:search_tech}

YAQQ can be used to compare two hard-coded gate sets based on the data set and decomposition method, as well as to decompose a unitary with a given gate set. 
This mode of operation can generate insights into already known gate sets or create a compiler on new data once a gate set is found.
However, the main utility of YAQQ is in generating a novel complementary gate set based on the data set and decomposition method.

While the Hilbert space of the data set grows exponentially, the space of the gate set is also continuous.
We consider various heuristics to explore this space intelligently.
The search happens in two steps.
Firstly, a gate set is defined in a parameterized manner.
Secondly, the parameters are adjusted based on the cost function.

In the second step, we use two techniques: either a stochastic search over the parameter space, which is denoted as RS or an optimization routine (e.g., SciPy), which encapsulates the cost function for a parametric gate set definition with a specific decomposition and data set.
There are various options for global optimization (e.g., Brute) or unconstrained minimization of multivariate scalar functions (e.g., Nelder-Mead, Powell, L-BFGS-B, COBYLA, SLSQP, etc.).
Our experiments found that the COBYLA optimizer~(CO) achieves faster and better convergence. 
This option is denoted as CO($\vec{p}$) where $\vec{p}$ are the vector of parameters defining the settings of the optimization.

\subsection{Novel gate set ansatz}\label{s2p5}


The gate set definition is an important step in pruning the search space.
It is well known that $2$-local gates are universal for quantum computation~\cite{divincenzo1995two}.
Circuits having gates with higher order gates can be decomposed, e.g., via the QSD.
Also, from the engineering perspective, almost all quantum processors have a modular gate set of one and two-qubit native gates.
There are some efforts to demonstrate $3$-qubit gates but with low fidelity.
In this work, we search over gate sets with $1$ and $2$-qubit gates.
Depending on the type of the gate, the search method can either be parametric or random.
Additionally, some standard constant gates can be part of the gate set but are not optimized.
YAQQ provides some built-in gates, listed in Table~\ref{gates} in Appendix~\ref{a2}.

A general n-qubit gate has $2^{2n}-1$ parameters; thus, a single qubit gate has $3$ free parameters while a two-qubit gate has $15$ free parameters.
A general one qubit gate can be specified using the P1 gate (equivalent to IBM's U3 gate) as:
\begin{equation}
\text{P1}_{\vec{a}} = \text{P1}_{a_1,a_2,a_3} = \begin{bmatrix}
cos(\frac{a_1}{2}) & -e^{ia_3}sin(\frac{a_1}{2})\\
e^{ia_2}sin(\frac{a_1}{2}) & e^{i(a_2+a_3)}cos(\frac{a_1}{2})
\end{bmatrix}
\label{eq:u3}
\end{equation}
For $2$-qubit gate, the $15$ parameters can be realized as a non-local unitary NL2 on the Weyl chamber with 3 coordinates sandwiched by 4 local 1-qubit gate P1 on each of the qubit before and after the NL2, thus, $3 + 4*3$.
The NL2 gate (often referred to as 2-qubit canonical gate) is defined as:
\begin{equation}
\text{NL2}_{\vec{t}} = \text{NL2}_{t_x,t_y,t_z} =  \exp\Biggl(-i \dfrac{\pi}{2} \bigg( t_x \begin{bmatrix}
0 & 0 & 0 & +1\\
0 & 0 & +1 & 0\\
0 & +1 & 0 & 0\\
+1 & 0 & 0 & 0
\end{bmatrix}_{X \otimes X} \hspace{-1.5em} + t_y \begin{bmatrix}
0 & 0 & 0 & -1\\
0 & 0 & +1 & 0\\
0 & +1 & 0 & 0\\
-1 & 0 & 0 & 0
\end{bmatrix}_{Y \otimes Y} \hspace{-1.5em} + t_z \begin{bmatrix}
+1 & 0 & 0 & 0\\
0 & -1 & 0 & 0\\
0 & 0 & -1 & 0\\
0 & 0 & 0 & +1
\end{bmatrix}_{Z \otimes Z}  
     \hspace{-1.5em}\bigg)\Biggl)
\label{eq:nl2}
\end{equation}

All other gates can be defined in terms of these two generic gates.
For an explicit definition of these gates, refer to \cite{crooks2020gates}.
The user can specify a desirable gate set with these options, e.g., \{R1, R1, CX2\}, or \{P1, T1, SPE2\}.
Note that we do not explicitly check if the gate set is universal. 
Thus, YAQQ can also be used for QEC applications where operations often belong to the non-universal Clifford group.
The total number of parameters is determined when the gate set is specified.
Thereafter, based on the chosen stochastic search or optimization technique option, the novel complementary gate set is generated.

\section{YAQQ software framework} \label{section:yaqq_sw}


Our open-sourced \texttt{qiskit} implementation of the novelty search on quantum gate sets is called Yet Another Quantum Quantizer~(YAQQ).
It is implemented in Python and available on the Python Package Index (PyPI)~\cite{yaqqpip}.
The package is designed to be intuitive to install and use for researchers without significant software development background.
For various mathematical operations and visualization, it depends on the Python libraries of \texttt{numpy}, \texttt{scipy}, \texttt{qutip}, \texttt{astropy}, \texttt{weylchamber}, \texttt{matplotlib}, and \texttt{tqdm}.
Thereafter, we present the workflow and the organization of the package modules.

\begin{figure}[htb]
         \centering
        \includegraphics[clip, trim=6cm 0cm 0cm 3cm,width=0.8\textwidth]{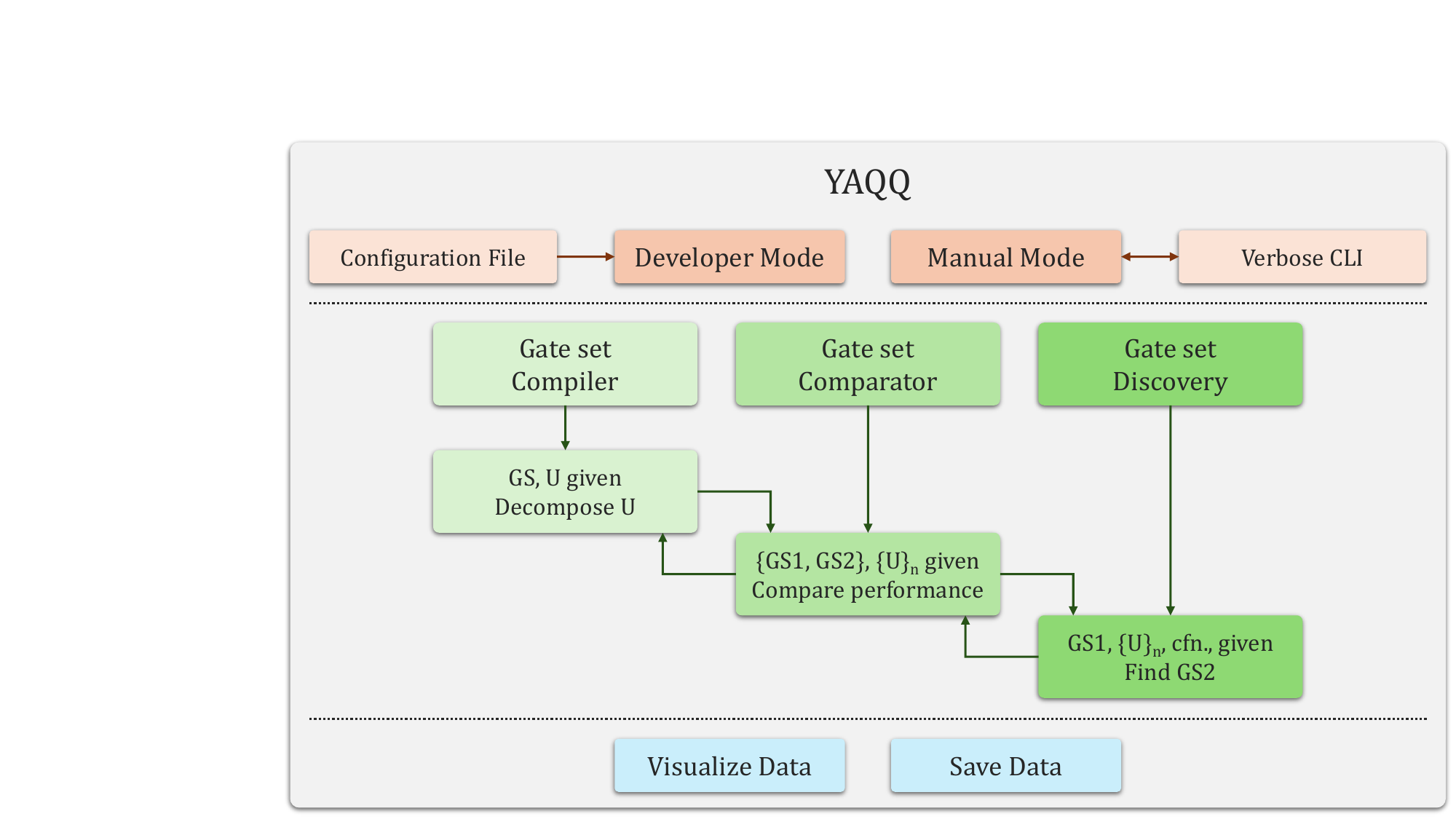}
         \caption{The workflow of YAQQ depicting the three usage levels, (i) gate set compiler, (ii) gate set comparator, and (iii) gate set discovery, and the nested dependencies between these levels. }         \label{fig:yaqq_blocks}
\end{figure}

The overall workflow of YAQQ is shown in Figure~\ref{fig:yaqq_blocks}.
YAQQ can be run in developer or manual mode.
The developer mode needs to specify a configuration file in the specific format that specifies all further details like the usage mode, data set, gate sets~(GS), hyperparameters, and the directory and filename for logging the results and plots.
In the manual mode, these above options can be configured via the command line interface~(CLI) based on the usage modes.
YAQQ has three usage levels: (i) gate set compiler, (ii) gate set comparator, and (iii) gate set discovery.
These three modes recursively depend on one another.
In the compiler mode, a gate set and a specific unitary $U$ are given, and YAQQ outputs the decomposed circuit, the associated process fidelity~(PF) of the approximation, and the circuit depth~(CD).
In the comparator mode, two gate sets, \{GS1, GS2\} and a set of unitaries \{U\}$_n$ is given.
The compiler mode is iteratively invoked for each unitary and both gate sets.
The statistics of the PF and CD form the performance comparison.
The discovery mode is the core novelty of this research.
In this mode, based on a defined cost function, a new gate set GS2 is iteratively refined for the data set, and another predefined gate set GS1.
This mode, in turn, invokes the comparator iteratively to compare a candidate GS2 against GS1.
YAQQ offers various options for visualizing the data set (for $1$ and $2$ qubits) and the results of the comparison between two gate sets for the chosen data set.
The plots, as well as the raw data, can be stored for reproducibility.

\begin{figure}[!ht]
         \centering
        \includegraphics[clip, trim=2.2cm 0cm 1.8cm 0cm,width=\textwidth]{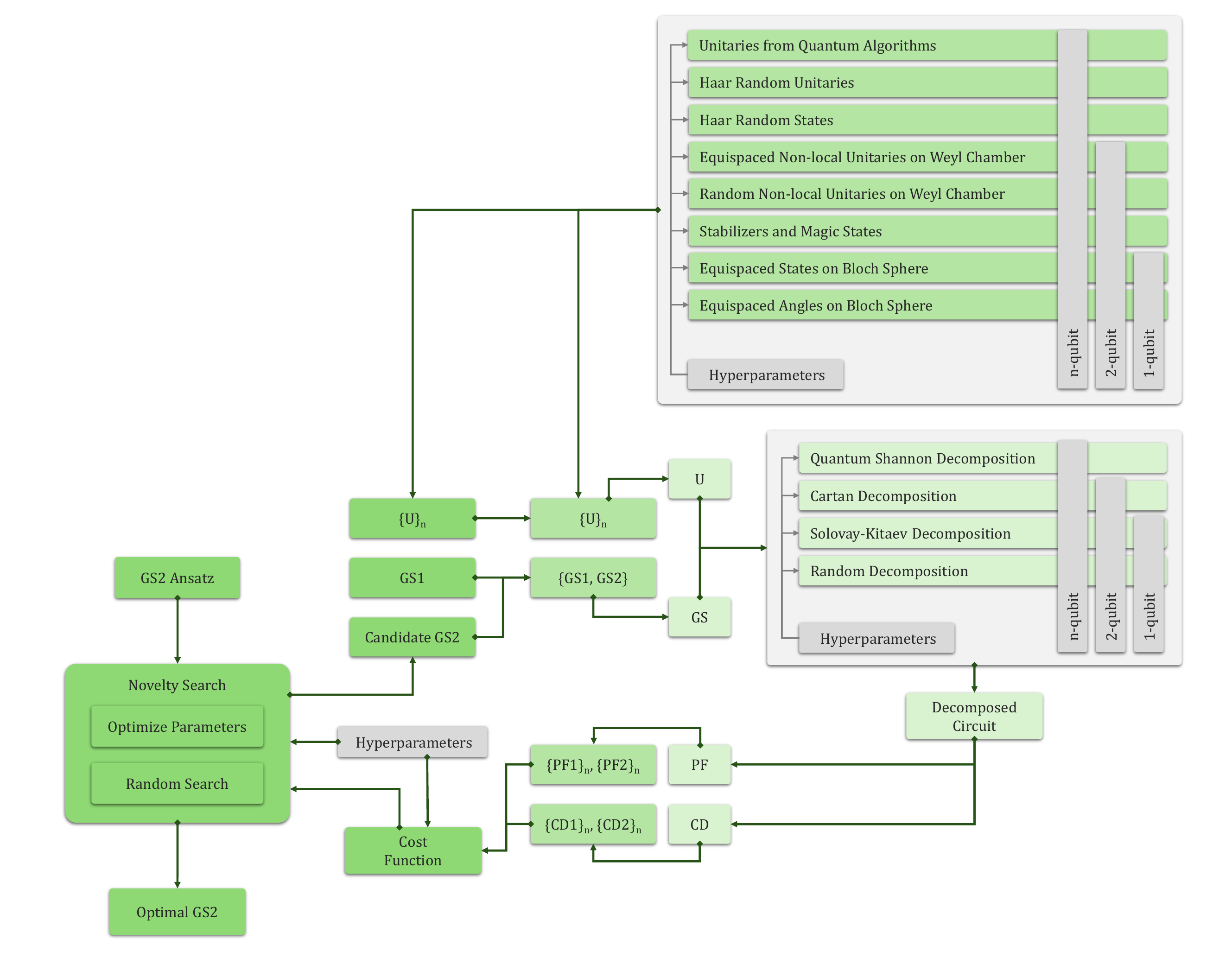}
         \caption{Module dependencies among the three usage levels. The innermost (light green) compiler modules decompose a single unitary based on a given gate set. The middle comparator level (green) iteratively uses the compiler for a specific dataset and two given gate sets, and compares the statistics of process fidelity and circuit depth. The outer discovery level (darker green) uses the comparator level iteratively to test a candidate gate set suggested by the novelty search based on the ansatz and assesses via a tunable cost function. A new optimal gate set with respect to the given gate set for the chosen data set and cost function is discovered in the process.
         }         \label{fig:yaqq_blocks_fine}
\end{figure}


A detailed organization of the module dependencies among the three usage levels is shown in Figure~\ref{fig:yaqq_blocks_fine}.
The innermost (light green) compiler modules decompose a single unitary based on a given gate set. 
Various decomposition options are available based on the size of the unitary.
The decomposition algorithms have associated hyperparameters like precision and number of trials that can be changed from the default value by the user.
The middle comparator level (green) iteratively uses the compiler for a specific dataset and two given gate sets and compares the statistics of process fidelity and circuit depth. 
The different dataset options, as discussed previously, are available based on the number of qubits.
The user also specifies the associated hyperparameters like dataset size or resolution.
The outer discovery level (darker green) uses the comparator level iteratively to test a candidate gate set suggested by the novelty search based on the ansatz and assesses via a tunable cost function. 
The hyperparameters at this level define the weights in the cost function, the optimization algorithm to use, and associated parameters.
A novel optimal gate set, conditioned on the given gate set, the chosen data set, and the cost function, is discovered in the process.

\section{Results} \label{section:results}


In this section, we present some benchmarks selected to evaluate YAQQ.
The sub-component configurations of these experiments are carefully designed to demonstrate specific features of YAQQ.
These designs and corresponding results are presented in the following sub-sections.
    
\subsection{Experiments with random datasets}

This section lists the experiments we conducted on YAQQ and the target insights that motivated these choices.
These proof-of-concept experiments were conducted on a dataset size of 10 random unitaries of the target dimension.
The GS1 is set of \{H1,T1\} for dimension 1 and \{H1,T1,CX2\} for dimension 2 or more.
This choice is motivated by the ubiquity of these gates as the standard gates when defining quantum algorithms and fault-tolerant quantum computation.
Note that for the SKT, it is implicit that the daggers are also considered in the gate set so that the group is closed under inversion.
These experiments are listed in Table~\ref{tab:exps}.
The first digit of the experiment ID (ExpID) denotes the dimension of the unitaries.

\begin{table}[!ht]
\centering
\caption{Experiments to evaluate novel gate set search}
\label{tab:exps}
\begin{tabular}{ |c|ccccc| } 
    \hline
    ExpID & GS2 & Dcmp. GS1 & Dcmp. GS2 & CFnWt. & Search \\ 
    \hline
    1.1 & \{P1$_{\vec{a}}$, P1$_{\vec{b}}$\} & SKT & SKT & [1,1,1,1,0] & CO \\ 
    1.2 & \{H1, T1, P1$_{\vec{a}}$\} & SKT & SKT & [1,1,1,1,0] & CO \\ 
    1.3 & \{P1$_{\vec{a}}$, P1$_{\vec{b}}$\} & RND & RND & [1,1,1,1,0] & CO \\ 
    1.4 & \{H1, T1, P1$_{\vec{a}}$\} & RND & RND & [1,1,1,1,0] & CO \\
    1.5 & \{P1$_{\vec{a}}$, P1$_{\vec{b}}$\} & RND & RND & $[50,1,1,1,0]$ & CO \\
    1.6 & \{H1, T1, P1$_{\vec{a}}$\} & RND & RND & $[50,1,1,1,0]$ & CO \\
    1.7 & \{H1, T1\} & SKT & RND & $[50,1,1,1,0]$ & Mode 2 \\
    1.8 & \{R1$_a$, R1$_b$\} & SKT & SKT & [1,1,1,1,0] & RS \\

    \hline
    
    2.1 & \{R1, P1$_{\vec{a}}$, SPE2$_{t_y}$\} & RND, KAK & RND, KAK & $[50,1,1,1,0]$ & CO \\   
    2.2 & \{P1$_{\vec{a}}$, P1$_{\vec{b}}$, SPE2$_{t_y}$\} & RND, KAK & RND, KAK & $[50,1,1,1,0]$ & CO \\ 
    2.3 & \{P1$_{\vec{a}}$, P1$_{\vec{b}}$, NL2$_{_{\vec{t}}}$\} & RND, KAK & RND, KAK & $[50,1,1,1,0]$ & CO \\ 
    
    \hline
    
    3.1 & \{P1$_{\vec{a}}$, P1$_{\vec{b}}$, SPE2$_{t_y}$\} & SKT, KAK, QSD & SKT, KAK, QSD & $[50,1,1,1,0]$ & CO \\   
    
    \hline
\end{tabular}
\end{table}





    
        
Each of these experiments aims to demonstrate some crucial features in YAQQ and quantum information structure in quantum gate sets.
Experiment 1.1 tries to find a complementary 1-qubit gate set composed of 2 discrete components using a parametric search over 6 parameters.
Experiment 1.2 aims to investigate the performance of over-specification of an already universal gate set.
It requires tuning 3 parameters for the additional P1 gate.
Experiments 1.3 and 1.4 compare the same task using the random decomposition method in YAQQ.
These demonstrate the usefulness of random decomposition for fast approximate decomposition with low depth cost when a slight loss in fidelity is permitted. 
Experiment 1.5 was designed to tune the cost hyperparameters.
The best results were found for the weight setting of [50,1,1,1,0], which were used for other experiments unless otherwise required.
Experiment 1.6 verifies these cost settings for the other GS2 specifications.
Experiment 1.7 used the compare gate set mode of YAQQ; no parameters were involved in either of the gate sets.
It is used to compare the two decomposition methods and tune the hyperparameters of RND (circuit length, trials, etc.) and SKT (recursion depth, basis approximation level, etc.).
Experiment 1.8 tests the Random search feature of YAQQ using 2 random gates, which, in principle, can find a universal set.
Experiments 2.1, 2.2, and 2.3 demonstrate YAQQ on 2 qubits.
The gate sets are capable of universality since NL2 or SPE2 can converge to CX2 and P1, or R1 can converge to H1 and T1.
The R1 gate is randomly fixed at the start, so, Experiment 2.1 is a 3+1 parameter optimization.
Experiment 2.2 is a 7-parameter optimization, while Experiment 2.3 is a 9-parameter optimization.
The Cartan decomposition method is also verified.
Note, the Cartan decomposition is not guaranteed to be exact if the NL2 is not a SPE2, however, YAQQ used that to find the closest approximation based on the fidelity.
Experiment 3.1 demonstrates the Quantum Shannon Decomposition for 2+ qubits using 7 parameters.
This is the most general setting where unitaries from runs of algorithms can be used as datasets to find gate sets that work best.
However, this can also be used for specific experiments, such as decomposing the Toffoli gate (just 1 unitary) while trading off fidelity and depth for various gate sets.

In the rest of this section, we provide a rigorous numerical benchmark of YAQQ. 
We use it to find a novel gate set with shallower average decomposition depth and higher average fidelity compared to typical fault-tolerant quantum computing gate sets of \{Hadamard, T, CNOT\} on $1$-, $2$- and $3$-qubit target unitaries.
These unitaries are generated using various data set generation methods discussed in Section~\ref{sec:data_set}. For the sake of this paper, we focus on the experiments provided in Table~\ref{tab:exps} specifically designed to benchmark YAQQ in the task of novel gate search.
In this task, the YAQQ uses the cost function described in Equation~\ref{eq:cost_function}. Given the multi-parameter dependency inherent in the cost function, it is crucial to appropriately configure the hyperparameters to reveal the true potential of YAQQ.

Hence, we split the assessment of numerical benchmarking of YAQQ into three sub-components: (1) hyperparameter optimization, where we investigate the suitable value of the parameters under which the cost function performs optimally; (2) optimal benchmarking, where we utilize the optimal settings of the hyperparameters to find a novel gate set that decomposes $1$-qubit unitaries with shallower depth quantum circuits with higher fidelity compared to native gate sets on IBM quantum devices and (3) optimal scalability, where to show the efficiency with increasing number of qubits we benchmark YAQQ in finding novel gate set to decompose $2$-qubit unitaries.


\subsubsection{Hyperparameter optimization}

The crucial hyperparameters responsible for enhancing the performance of YAQQ are the weights of the cost function (Equation~\ref{eq:cost_function}), the number of parameterized gates in the novel gate set, and the search method we utilize to optimize the parameters of these gates. Another relatively less crucial hyperparameter that can affect the performance of YAQQ is the choice of the decomposition technique. In the following, we will discuss each of these parameters individually and their corresponding optimal setting to reveal the highest potential of YAQQ. We use the experiments $1.1$-$1.8$ (from Table~\ref{tab:exps}) to find the optimal hyperparameters.

\begin{figure}[!th]
    \centering
\includegraphics[scale=0.6]{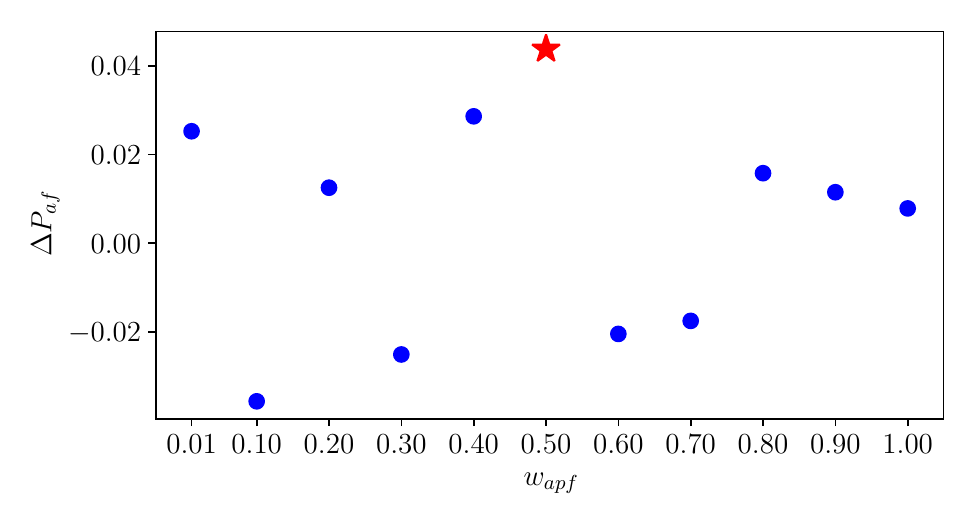}\caption{The optimal configuration of the cost function by adjusting the weights. The experiment is conducted on a dataset containing $1$-qubit Haar random unitaries. The novel gate set contains \{H1, T1\} and a parameterized rotation P1$_{\vec{a}}$, optimized using the SciPy search. The performance of the optimized gate set is then compared with the primitive novel gate set \{H1, T1\}. Using these two gate sets, we decompose the dataset using the SKT. We investigate the variation in the difference in average process fidelity ($\Delta \langle\text{PF}\rangle$, where $\langle\text{PF}\rangle$ is the average process fidelity for a specific gate set) in decomposing $10$ data points from the dataset with respect to the variation of $w_{apf}$ while keeping the other weights fixed. The more positive the difference is, the more we improve the \{H1, T1, P1$_{\vec{a}}$\} gate set compared to \{H1, T1\}. We observe the cost function performs optimally at $w_{apf}=50$, indicating the optimal configuration of the cost function [$w_{apf}$, $w_{npf}$, $w_{acd}$,  $w_{ncd}$, $w_{agf}$] = [50,1,1,1,0]. 
 }
    \label{fig:weight_choice}
\end{figure}

\paragraph{Adjusting the weights} 

$w_{apf}$, $w_{npf}$, $w_{acd}$,  $w_{ncd}$ and $w_{agf}$ are the five weights that need to be adjusted in YAQQ to optimize the cost function (Equation~\ref{eq:cost_function}) to find the optimal novel gate sets. For the sake of simplicity, we denote them as the list of weights where $W$ = [$w_{apf}$, $w_{npf}$, $w_{acd}$,  $w_{ncd}$, $w_{agf}$]. For the sake of further simplicity, we choose the list as [$w_{apf}\geq1$, $1$, $1$, $1$, $0$], which means we are encouraging the YAQQ to find a novel gate set that improves the average process fidelity compared to the novel gate set (in this case is \{H1, T1, TD1\}) while turning off the quantum engineering part (which is relevant while dealing with real quantum devices) by setting $w_{agf}=0$. 

In Figure~\ref{fig:weight_choice}, we plot the difference of process fidelity averaged over $10$, $1$-qubit Haar, random unitaries using the novel gate set \{H1, T1, P1$_{\vec{a}}$\} and \{H1, T1\} with respect to the weight $w_{apf}$. 
We observe that the YAQQ performs optimally, denoting the optimal configuration of the cost function when we set $w_{apf}=0.5$, keeping other weights fixed.
\begin{table}[!ht]
\centering
\begin{tabular}{ |c|c|c|c| } 
\hline
Search & Gateset & $\langle\text{PF}\rangle$ & $\langle\text{CD}\rangle$ \\
\hline
\multirow{4}{4em}{Random} & \{H1, T1\} & 0.9432 & 255.45\\ 
& \{P1$_{\vec{a}}$, P1$_{\vec{b}}$\} & \cellcolor{red!15}0.9331 & \cellcolor{green!15}242.4\\ \cline{2-4}
& \{H1, T1\} & 0.9389 & 284.7\\
& \{H1, T1, P1$_{\vec{a}}$\} & \cellcolor{red!15}0.9366 & \cellcolor{green!15}222.55\\
\hline
\multirow{4}{4em}{SciPy} & \{H1, T1\} & 0.9167 & 258.65 \\ 
& \{P1$_{\vec{a}}$, P1$_{\vec{b}}$\} & \cellcolor{green!15}0.9324 & \cellcolor{green!15}246.9 \\
\cline{2-4} 
& \{H1, T1\} & 0.9426 & 271.3 \\
& \{H1, T1, P1$_{\vec{a}}$\} & \cellcolor{red!15}0.9339 & \cellcolor{red!15}303.9 \\
\hline
\end{tabular}
\quad
\begin{tabular}{ |c|c|c|c| } 
\hline
Search & Gateset & $\langle\text{PF}\rangle$ & $\langle\text{CD}\rangle$ \\
\hline
\multirow{4}{4em}{Random} & \{H1, T1\} & 0.9161 & 239.4\\ 
& \{P1$_{\vec{a}}$, P1$_{\vec{b}}$\} & \cellcolor{green!15}0.9469 & \cellcolor{red!15}269.2 \\ \cline{2-4}
& \{H1, T1\} & 0.9310 & 221.65\\
& \{H1, T1, P1$_{\vec{a}}$\} & \cellcolor{red!15}0.9193 & \cellcolor{red!15}264.65\\
\hline
\multirow{4}{4em}{SciPy} & \{H1, T1\} & 0.9285 & 260.55 \\ 
& \{P1$_{\vec{a}}$, P1$_{\vec{b}}$\} & \cellcolor{green!15}0.9489 & \cellcolor{green!15}249.4 \\
\cline{2-4} 
& \{H1, T1\} & 0.9305 & 218.7 \\
& \{H1, T1, P1$_{\vec{a}}$\} & \cellcolor{green!15}0.9468 & \cellcolor{red!15}337.5 \\
\hline
\end{tabular}
\caption{Comparing the performance of the weight distribution [1,1,1,1,0] (left-hand side table) and [50,1,1,1,0] (right-hand side table) using two kinds of novels gate sets and search methods. We compare the primitive native gate set \{H1, T1\} with novel parameterized ansatz gate sets \{H1, T1, P1$_{\vec{a}}$\} and \{P1$_{\vec{a}}$, P1$_{\vec{b}}$\}. The YAQQ then uses either a random search or a SciPy optimizer method to find the optimal parameters of the gate by optimizing $\vec{a}$ (and $\vec{b}$). From the weight distribution, it is evident that we are putting more stress on the average process fidelity ($\langle\text{PF}\rangle$) than the average circuit depth ($\langle\text{CD}\rangle$) of the decomposition achieved by novel gate sets, hence increasing the depth of decomposition for a higher fidelity is an acceptable trade-off. Hence, for a fixed optimization method, the [50,1,1,1,0] outperforms [1,1,1,1,0] in terms of providing us with a better average PF, further motivating the fact that the optimal cost function configuration can be achieved by fixing the weight [50,1,1,1,0].}
\label{tab:11110_and_501110_comparison}
\end{table}

In Figure~\ref{fig:weight_choice}, the weights [1,1,1,1,0] and [50,1,1,1,0] closely compete with each other in terms of providing us with the novel gate set that decomposes a Haar random unitary with very high fidelity. Hence, in Table~\ref{tab:11110_and_501110_comparison}, we benchmark the performance of these two cost functions corresponding to the weights. As we emphasize more the average process fidelity ($\langle\text{PF}\rangle$) than the average circuit depth ($\langle\text{CD}\rangle$), an enhancement in PF for a deeper unitary decomposition is an acceptable trade-off. For a fixed search method and gate set ansatz, we see that the cost function with the weight distribution [50,1,1,1,0] outperforms the cost corresponding to the weight list [1,1,1,1,0].  This investigation further proves that the optimal choice of weights for the cost function is [50,1,1,1,0] irrespective of the novel gate set ansatz.

\paragraph{Preferred optimization method} The main two optimization methods YAQQ utilizes are based on random search and SciPy optimizers, which are elaborately discussed in Section~\ref{sec:search_tech}. For the optimal setting of the cost function with weight distribution  [50,1,1,1,0], in the right-hand side table of Table~\ref{tab:11110_and_501110_comparison} we can see that using random search method the novel gate set ansatz \{P1$_{\vec{a}}$, P1$_{\vec{b}}$\} can achieve a higher fidelity compared to the native gate set \{H1,T1\} but the \{H1, T1, P1$_{\vec{a}}$\} fails achieve it. Whereas when using the SciPy optimizer, we see that for \{H1, T1, P1$_{\vec{a}}$\} the optimizer can find the optimal value of $\vec{a}$ that returns a higher fidelity in the decomposition task compared to the native gate set \{H1, T1\}. Interestingly, the SciPy optimizer even enhances the performance of the \{P1$_{\vec{a}}$, P1$_{\vec{b}}$\} by providing us with a better average fidelity with a shallower depth decomposition, indicating the fact that SciPy optimizer is preferable as a search method in comparison to random search.

\paragraph{The choice of decomposition} In  Section~\ref{sec:decomp_tech} we discuss two ways to decompose a $1$-qubit unitary. The first technique corresponds to randomly decomposing the unitaries, whereas in the second case, we use Solovay-Kiteav (SKT) decomposition. 
\begin{table}[!htb]
\centering
\begin{tabular}{ |c|c|c|c| } 
\hline
Search & Gateset & $\langle\text{PF}\rangle$ & $\langle\text{CD}\rangle$ \\
\hline
\multirow{4}{4em}{Random} & \{H1, T1\} & 0.7479 & 41.5\\ 
& \{P1$_{\vec{a}}$, P1$_{\vec{b}}$\} & \cellcolor{green!15}0.8131 & \cellcolor{green!15}5.05 \\ \cline{2-4}
& \{H1, T1\} & 0.6537 & 42.8\\
& \{H1, T1, P1$_{\vec{a}}$\} & \cellcolor{green!15}0.9166 & \cellcolor{green!15}21.05\\
\hline
\multirow{4}{4em}{SciPy} & \{H1, T1\} & 0.7751 & 45.25 \\ 
& \{P1$_{\vec{a}}$, P1$_{\vec{b}}$\} & \cellcolor{green!15}0.9268 & \cellcolor{green!15}5.75 \\
\cline{2-4} 
& \{H1, T1\} & 0.7531 & 43.1 \\
& \{H1, T1, P1$_{\vec{a}}$\} & \cellcolor{green!15}0.9217 & \cellcolor{green!15}13.6 \\
\hline
\end{tabular}
\caption{The performance of the Solovay-Kiteav (SKT) decomposition investigated over two novel gate set ansatz and search methods. While using the SKT decomposition, we observe that it gives a very shallow decomposition of unitaries compared to random decomposition, irrespective of the gate set and optimization method. Furthermore, we observe a notable decrease in average process fidelity ($\langle\text{PF}\rangle$) using \{H1, T1\} gate set. But the same fidelity increases when the ansatz novel gate sets \{P1$_a$, P1$_b$\} and \{H1, T1, P1$_a$\} are utilized.}
\label{tab:skt_decomp}
\end{table}
In Table~\ref{tab:skt_decomp}, we utilize the SKT decomposition instead of a random one. The investigation demonstrates that SKT decomposition is depth-efficient compared to a random decomposition. Moreover, random search and the SciPy optimizers provide us with a higher average gate fidelity with nearly eight times less circuit depth for \{P1$_a$, P1$_b$\} ansatz gate set. Whereas in the case of \{H1, T1, P1$_a$\} we get a higher average fidelity but with nearly two (with random search) to three (with SciPy optimizers) times less circuit depth. This motivates us to use the SKT as the optimal decomposition for higher fidelity and a shorter depth unitary decomposition.

In summary, our initial investigation reveals that YAQQ achieves optimal performance under the following conditions: (1) setting the weights of the cost function (as defined in Equation~\ref{eq:cost_function}) to [50,1,1,1,0], (2) optimizing the parameters of the ansatz gate set using the scipy COBYLA optimizer, and (3) decomposing the unitaries using the Solovay-Kitaev decomposition method.

\paragraph{Novel gate set design}

The benchmarking of the results of YAQQ primarily focuses on the enhancement of average case fidelity of the ansatz gate set, giving little attention to the novelty of the gate set. 
\begin{figure}[!ht]
    \centering
    \includegraphics[scale = 0.5]{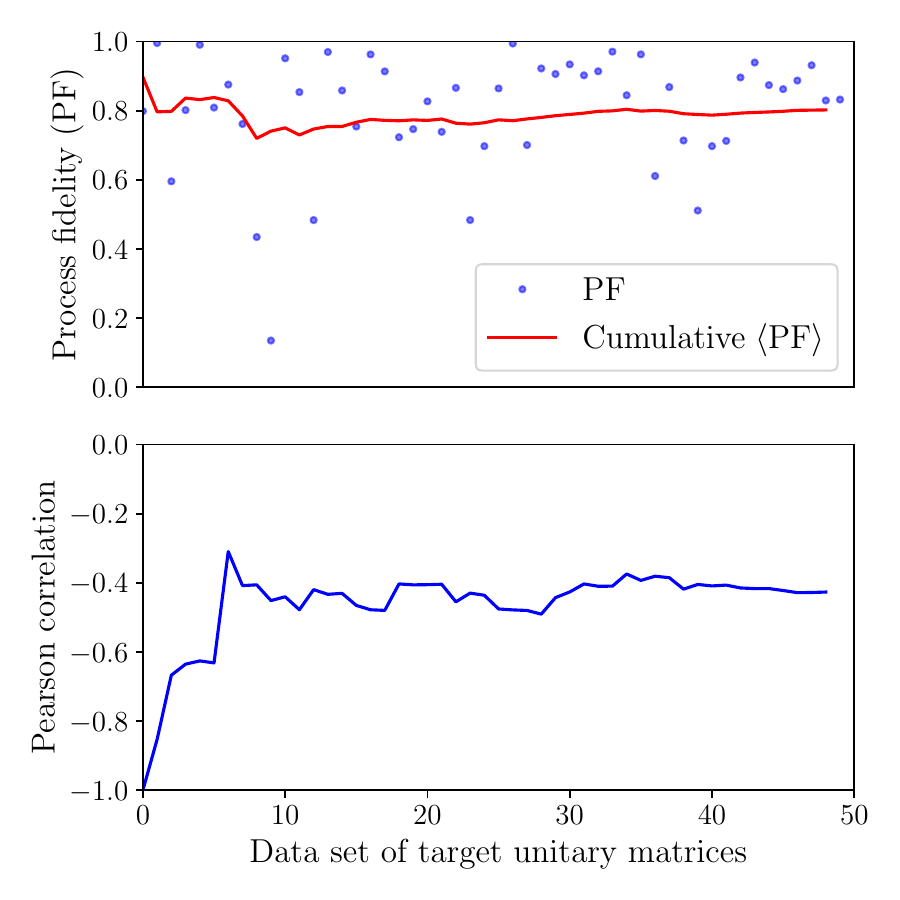}
    \caption{The novelty in the ansatz gate set \{P$1_{\vec{a}}$, P$2_{\vec{b}}$\}. The novelty is amplified by modifying the weight $w_{npf}$ in the cost function. $w_{npf}$ weights the novelty in the process fidelity, making the process fidelity (PF) achieved with the \{P$1_{\vec{a}}$, P$2_{\vec{b}}$\} anti-correlated to the PF achieved with \{H1, T1\}. The anti-correlation in process fidelity increases as we consider more data but saturates close to $-0.5$, which still corresponds to a strong anti-correlation between the PF of ansatz and the novel gate set.}
    \label{fig:novelty_in_gateset}
\end{figure}
In this section, we further benchmark the performance of YAQQ by giving equal focus on the novelty of the ansatz gate set while decomposing $50$, 1-qubit Haar random unitaries. The crucial factor that impacts the novelty of the ansatz gate set is choosing the cost function weights carefully. As the quantifier of the novelty, we utilize Pearson's correlation function ($P_{crr}$), which varies
between $-1$ to $1$ where if $-1\leq P_{crr} <0$, then it defines anti-correlation and $0> P_{crr}\geq1$ corresponds correlation between two given dataset. Novelty in the gate set corresponds to an anti-correlation between the process fidelity obtained through the ansatz gate set and the novel gate set.  For this experiment, \{P$1_{\vec{a}}$, P$2_{\vec{b}}$\} were chosen as the ansatz gate set and \{H1, T1\} as the novel gate set. 

We recall the weight $w_{npf}$ in the cost function impacts the novelty of the ansatz gate set. In Fig.~\ref{fig:novelty_in_gateset}, we set $w_{npf}=20$, which modifies the optimal setting of the weights to [50,20,1,1,0], and we can see that the average Pearson correlation of the process fidelity obtained by the gate sets strongly anti-correlates (i.e. $P_{crr}\leq0.5$) for $10$ Haar random unitaries. Afterward, as we increase the number of samples in unitary, the anti-correlation saturates close to $-0.5$.

\begin{figure}[!ht]
         \centering
        \includegraphics[scale=0.55]{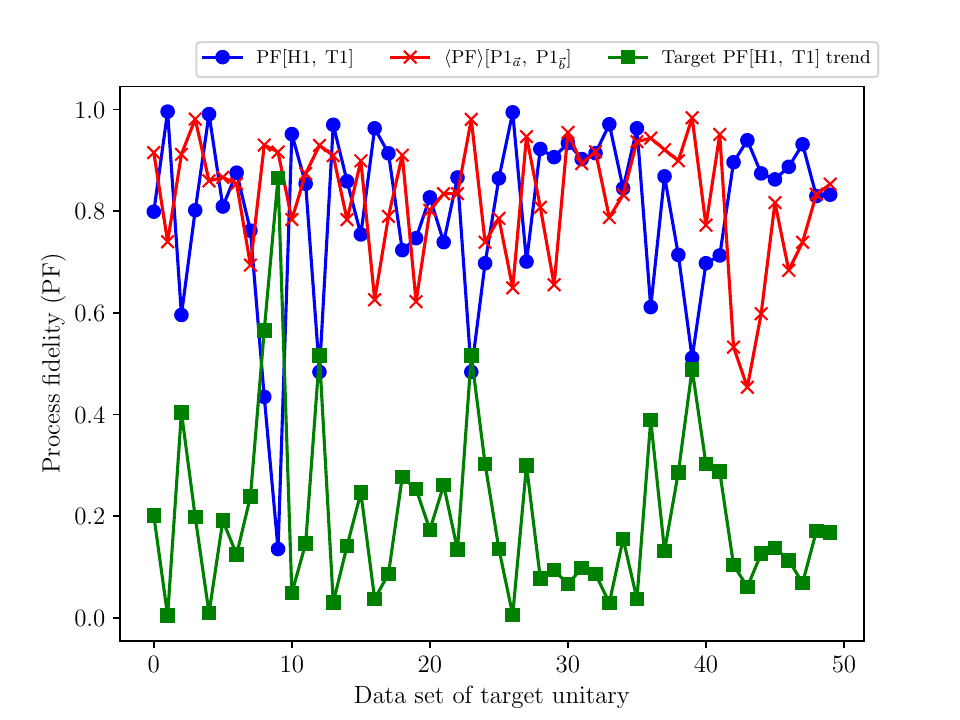}
         \caption{
         In this task, we use $50$, 1-qubit Haar random unitaries, which are decomposed using a novel gate set \{H1, T1\} and ansatz gate set \{P$1_{\vec{a}}$, P$2_{\vec{b}}$\} using SKT decomposition. The YAQQ optimizes the ansatz gate set using $500$ iterations of \texttt{COBYLA} optimizer. The process fidelity (PF) obtained by the optimized ansatz gate set is strongly anti-correlated to the PF we get from the novel gate set, representing the novelty in the ansatz gate set. The $\langle\text{PF}\rangle$ for the novel gate set is higher than the original.
         }
         \label{fig:novel_50_points}
\end{figure}

In Fig.~\ref{fig:novel_50_points}, we plot the process fidelity of decomposing $50$ random unitaries for the novel and the ansatz gate set with the weight distribution of cost function [50,20,1,1,0]. The results show that the trend in the variation of the process fidelity in the ansatz gate set is anti-correlated to the process fidelity obtained through the novel gate set. Moreover, manipulating the cost function induces a strong novelty in the optimized ansatz gate set, but the average fidelity also increased compared to the novel gate set.

After optimizing the ansatz gate set \{P$1_{\vec{a}}$, P$2_{\vec{b}}$\} using $500$ iterations of \texttt{COBYLA} optimizer we obtain the novel gate set as follows:
$$\text{P1}^{nov.50}_{\vec{a}} = \begin{bmatrix}
+0.95695777+0.j & -0.25428694-0.13989277j \\
+0.20772632+0.20268599j & +0.27797849+0.91569434j
\end{bmatrix},$$

$$\text{P1}^{nov.50}_{\vec{b}} = \begin{bmatrix}
-0.23161363+0.j & -0.97275633+0.01001202j \\
+0.96339857-0.13497524j & -0.22903051+0.03449493j
\end{bmatrix}.$$


\subsubsection{Benchmarking gate set optimality}

Utilizing the optimal setting from the previous section, we benchmark the YAQQ's performance in designing a gate set that provides an optimal decomposition to Haar random unitaries. Throughout the experiments, we consider \{$\text{P1}_{\vec{a}}$, $\text{P1}_{\vec{b}}$\} as the ansatz gate set where the parameters $\vec{a}$ and $\vec{b}$ are optimized using $1000$ iterations of COBYLA optimizer (CO). 

\begin{figure}[!ht]
         \centering
         \includegraphics[width=\textwidth]{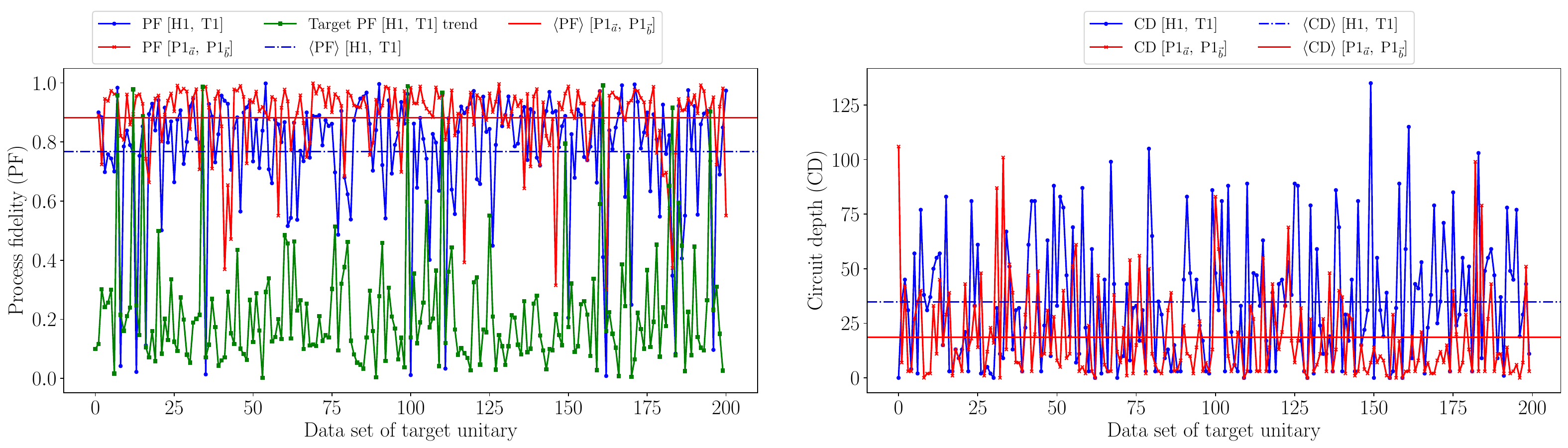}
         \caption{YAQQ results simulated on open source quantum software \texttt{qiskit}. The results are gathered while decomposing $200$ Haar random unitaries using the novel gate set \{H1, T1\} and the ansatz gate set \{P1$_a$, P1$_b$\}. The first thing to be noted is that using the ansatz gate set, we can decompose the unitaries with higher average fidelity than the average of the novel gate set. Moreover, the decomposition depth of unitaries is shallower using the ansatz gate set than the novel gate set.
         }
         \label{fig:benchmark_200_points}
\end{figure}

In Figure~\ref{fig:benchmark_200_points}, we consider $200$ Haar random unitaries where we show that the average process fidelity and the depth with the ansatz gate set outperforms the novel gate set. In the optimal configuration of the cost function, the weight $w_{npf}$, which is responsible for deciding the novelty of the YAQQ, is set to $1$. That is why even though we get an increase in average process fidelity with a shallower decomposition using the ansatz gate set, the gate set lacks novelty. Hence, motivated by this observation, in the next experiment, we modify the weights of the optimal cost function by turning $w_{npf}$ as a variable. So, the weights of the cost function become [$50$,$w_{npf}$,$1$,$1$,$0$]. As the $w_{npf}$ increases, we expect to observe an increase in anti-correlation between the trend in average process fidelity obtained using the novel and the ansatz gate set. To measure anti-correlation, we utilize Pearson's correlation coefficient~\cite{pearson1895vii}.

$$\text{P1}^{opt.200}_{\vec{a}} = \begin{bmatrix}
-0.78687181+0.j & -0.04577325-0.61541657j \\
+0.03316088-0.61622488j & -0.7867068 +0.01611391j
\end{bmatrix}$$

$$\text{P1}^{opt.200}_{\vec{b}} = \begin{bmatrix}
+0.8915669 +0.j & +0.22308591-0.39413341j \\
+0.24090997-0.38349818j & +0.42340431+0.78461476j
\end{bmatrix}$$

\begin{figure}[!htb]
     \centering
     \begin{subfigure}[b]{0.24\textwidth}
         \centering
         \includegraphics[clip, trim=0cm -1cm 0cm 0cm,width=0.99\textwidth]{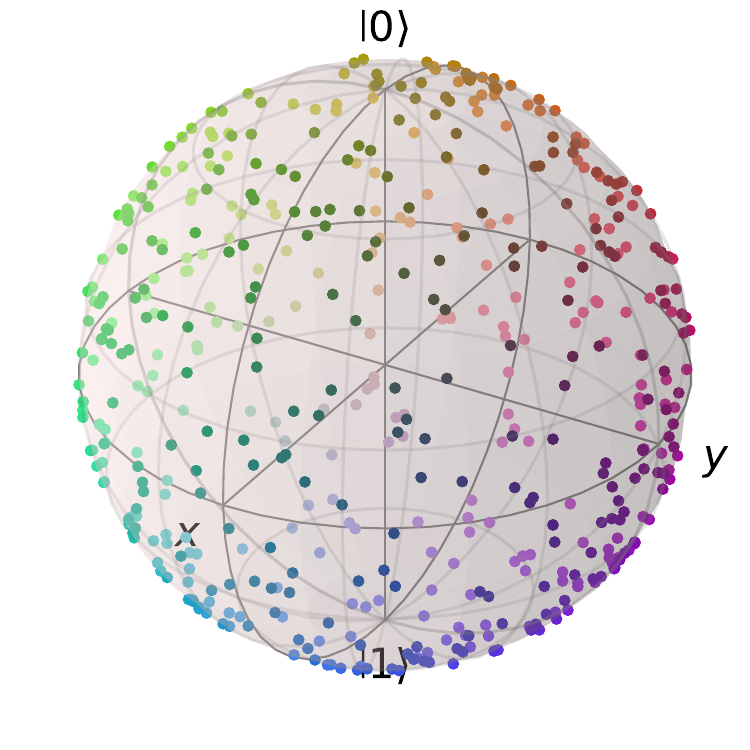}
         \caption{\small{Visualization of $500$ Haar random unitaries}.}
         \label{fig:1q500ds}
     \end{subfigure}
     \hfill
     \begin{subfigure}[b]{0.75\textwidth}
         \centering
         \includegraphics[scale=0.6]{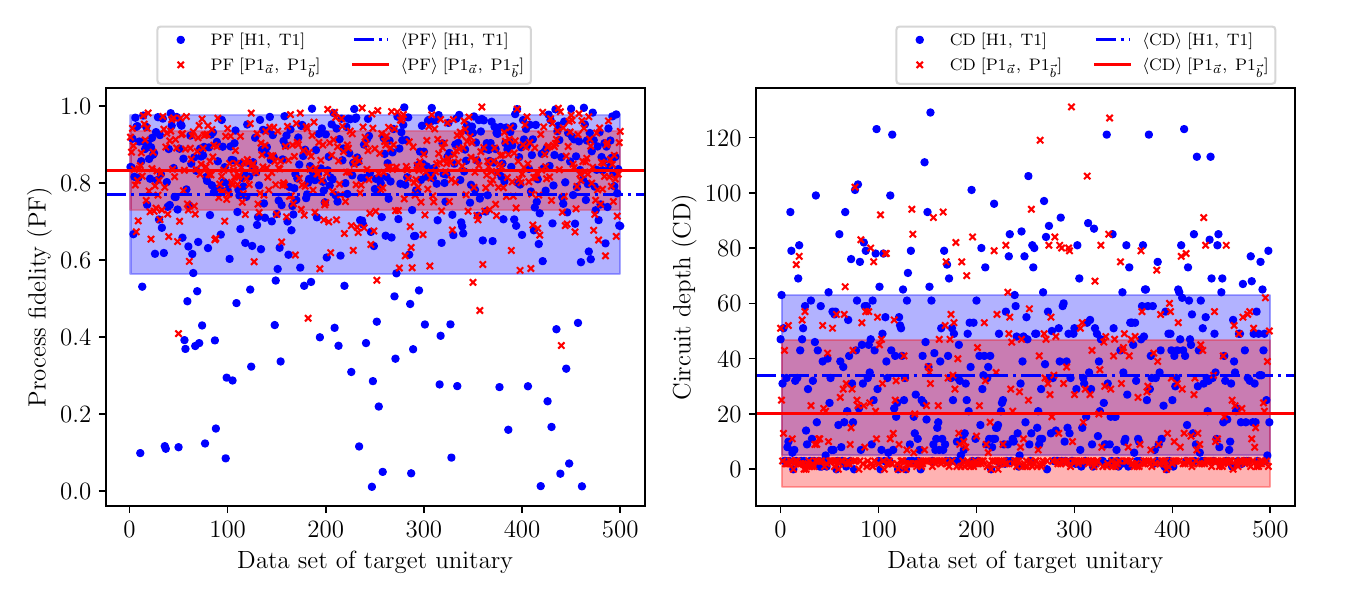}
         \caption{The comparison of the \{H1, T1\} and the ansatz gate set \{P1$_a$, P1$_b$\}.}
         \label{fig:1q500cmp}
     \end{subfigure}
        \caption{Comparison of the performance of YAQQ in decomposing $500$ Haar random unitaries with \{H1, T1\} and the ansatz gate set \{P1$_{\vec{a}}$, P1$_{\vec{b}}$\}. In this experiment, 
        we choose the optimal setting of the YAQQ where the Haar random unitaries are decomposed using SKT decomposition, the weights of the cost function are set to [50,1,1,1,0], and the parameters of the ansatz gate set are optimized using SciPy COBYLA optimizer. As the performance quantifier, we investigate the process fidelity (PF) and the circuit depth (CD). In figure (b), we can see that the $\langle\text{PF}\rangle$ of decomposing $500$ unitaries with the novel ansatz \{P1$_{\vec{a}}$, P1$_{\vec{b}}$\} is more accurate compared to the \{H1, T1\} gate set. Furthermore, the ansatz gate set accomplishes this task using a shallower circuit.}
        \label{fig:benchmark_500_points}
\end{figure}

In Figure~\ref{fig:benchmark_500_points}, we then extend the results to $500$ Haar random unitaries and show that the YAQQ can fine-tune the \{P1, P1\} gate set for fidelity.
The gates found are:
$$\text{P1}^{opt.500}_{\vec{a}} = \begin{bmatrix}
-0.99891178+0.j & -0.01683013-0.04349723j \\
-0.0406026 +0.02294975j & +0.7722138 +0.63364862j
\end{bmatrix}$$
$$\text{P1}^{opt.500}_{\vec{b}} = \begin{bmatrix}
+0.54683177+0.j & -0.52979855-0.64829662j \\
+0.8231603 +0.15291220j & +0.26287606+0.47950095j
\end{bmatrix}$$


\subsubsection{Optimal scalability}

\begin{figure}[!ht]
         \centering
         \includegraphics[scale=0.6]{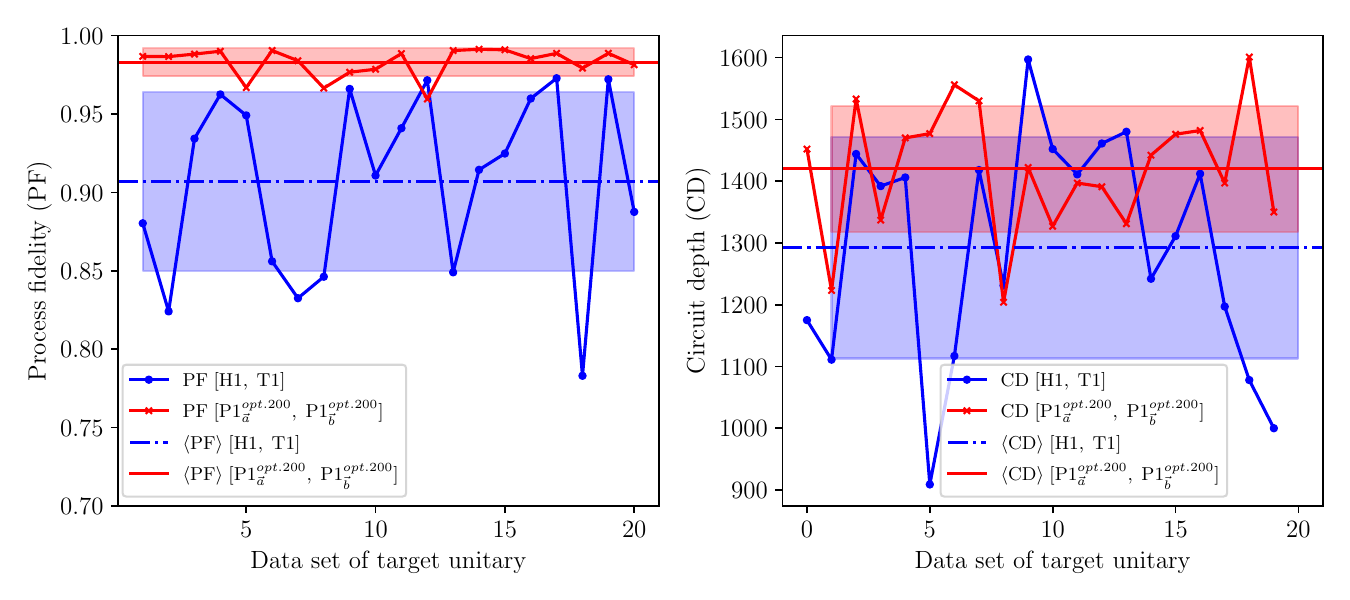}
         \caption{The original and optimal novel gate sets from the experiment presented in Fig.~\ref{fig:benchmark_200_points} are extended with the CX2 gate. This allows the decomposing of n-qubit unitaries. Here, the performance of $20$ Haar random $2$-qubit unitary matrices is decomposed using the KAK and SKT. This demonstrates the scalability of the YAQQ by generalizing results to a larger Hilbert space.
         }
         \label{fig:novel_10_points_2q}
\end{figure}

The optimal gate set on one qubit data set can now be augmented to a general 2-qubit gate set by adding a special perfect entangler.
We extend both the original gate set \{H1, T1\} and \{$\text{P1}^{opt.200}_{\vec{a}}$, $\text{P1}^{opt.200}_{\vec{b}}$\} gate set by adding CX2.
The comparison of the performance of these two gate sets is shown in Figure~\ref{fig:novel_10_points_2q}.
For this experiment, the optimal P1 gates are supplied to YAQQ via the F1 (load gate from file) option, based on the saved gates from the previous experiment of Figure~\ref{fig:benchmark_200_points}.
We demonstrated the scalability of the novel gate set for multi-qubit decomposition.
The two-qubit decomposition uses the KAK decomposition.
Similarly, this gate set can also be used to decompose $n$-qubit unitaries via the QSD option.

\subsection{Experiments with application-specific datasets} \label{sec:apps}

This section presents three exemplary applications where YAQQ can be applied.
We also discuss some related works and approaches used in other toolsets for these applications.


\subsubsection{Comparison of quantum error correction codes} \label{app:qec}

Transversality is an important feature of FTQC.
Transversal logical gates are those for which an error-correcting code can achieve a transformation on a logical qubit by applying that gate to each of the physical qubits~\cite{lidar2013quantum}.
For example, in a $7$-qubit Steane code, logical Hadamard can be performed by applying Hadamard on each of the $7$ physical qubits
This feature makes these gates the simplest to implement because they are automatically fault-tolerant, i.e., if an error occurs on one physical qubit, no action can propagate it onto a different physical qubit in the same code because no two qubits in the block every interact. 
These logical gates are typically also short gate sequences, making the corresponding fault-tolerant threshold particularly high.
However, the Eastin-Knill theorem~\cite{eastin2009restrictions} states that a universal gate set cannot be transversal in any quantum error correction~(QEC) code.
Comparing transversal logical gates to the fault-tolerance capabilities of a QEC code is an active research field~\cite{howard2017application} that can benefit from empirical experiments on YAQQ.

\begin{figure}[!htb]
     \centering
     \begin{subfigure}[b]{0.29\textwidth}
         \centering
         \includegraphics[clip, trim=0cm 0cm 0cm 0cm,width=0.99\textwidth]{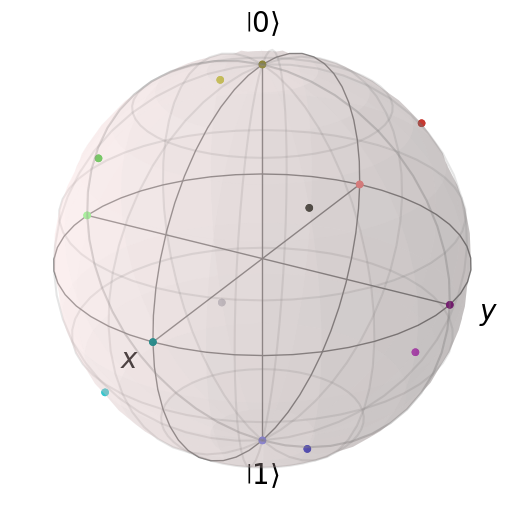}
         \caption{1 qubit dataset of 6 stabilizers and 8 magic states}
         \label{fig:qecc1}
     \end{subfigure}
     \hfill
     \begin{subfigure}[b]{0.69\textwidth}
         \centering
         \includegraphics[clip, trim=3.5cm 1cm 4.4cm 2.2cm,width=0.99\textwidth]{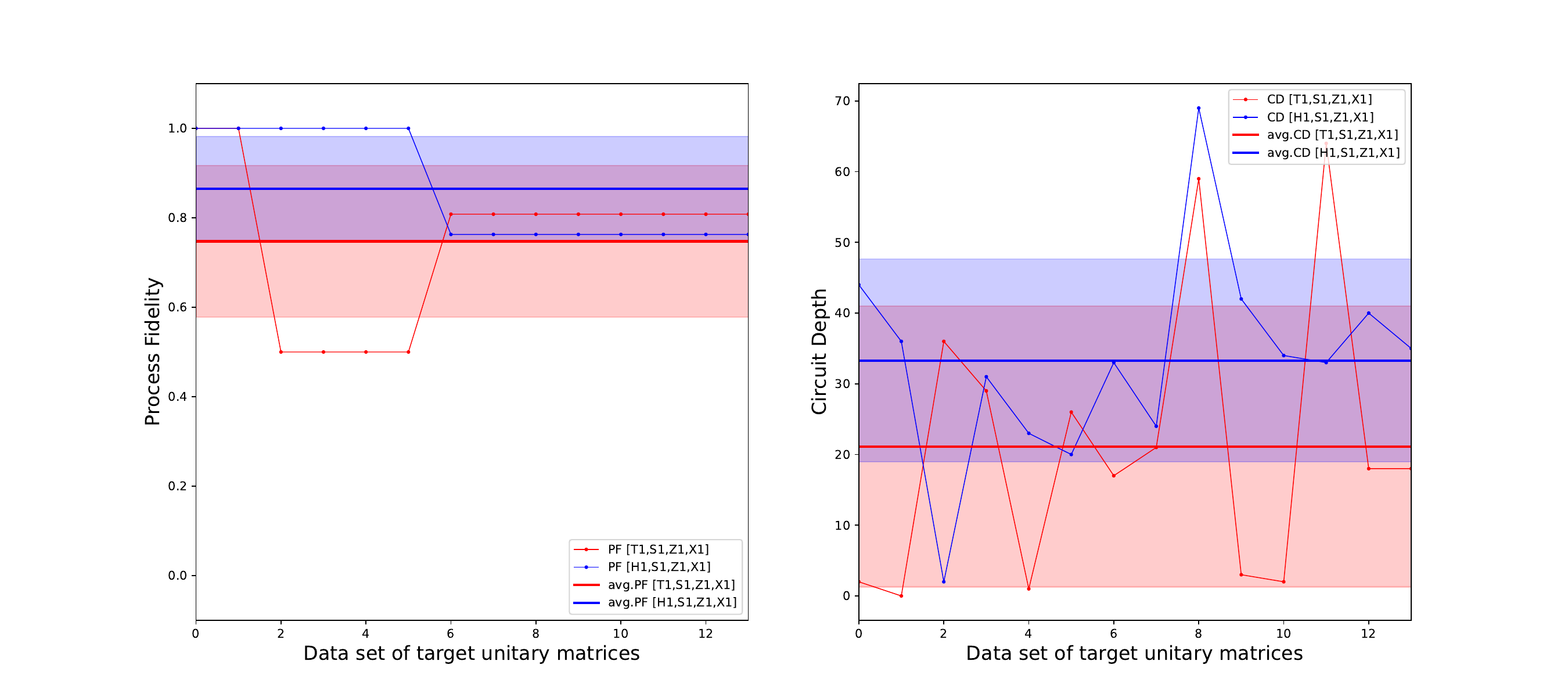}
         \caption{Comparison result}
         \label{fig:qecc2}
     \end{subfigure}
        \caption{Comparison of transversal gate sets in quantum Reed-Muller and Steane codes. We infer that the fidelities of magic states (data points $6-13$) are better with the Reed-Mueller code (in red), while the fidelities of cardinal states (data points $0-5$) in the X and Y direction are difficult without an H1 gate. Conversely, the stabilizer states are trivial for the Steane code (in blue) while failing to reach the magic states.}
        \label{fig:qecc}
\end{figure}

In an experiment to compare transversal logical gate sets, we used the YAQQ's compare functionality.
We specifically compared the Reed-Mueller code, having the transversal gate set \{T1, X1, S1, Z1, CZ2\}, with the Steane code, having the transversal gate set \{H1, X1, S1, Z1, CX2\}~\cite{anderson2014fault}. 
The dataset was specifically designed for the $1$-qubit quantum states composed of $6$ stabilizers and $8$ magic states, as shown in Figure~\ref{fig:qecc1}.
These are typically considered the easiest and hardest states to represent in QEC.
However, as a tradeoff, codes with easy preparation of Clifford states (e.g., in Steane or CSS codes) don't have the easy preparation of magic gates like T1 (e.g., in Reed-Mueller).
This tradeoff is shown in Figure~\ref{fig:qecc2}, where the fidelity of magic states is better with the Reed-Mueller (in red) code, while the fidelity of cardinal states in the X and Y direction is difficult without an H1 gate.
Conversely, the stabilizer states are trivial for the Steane code, while they fail for the magic states.
Based on experiments, we found the Reed-Mueller code to be more amenable to the Solovay-Kitaev decomposition, and conversely, the Steane code performed better for random decomposition.
The raw experiment data is available on the YAQQ Git repo.


As a generalization of the QEC application, YAQQ can be used to empirically demonstrate weak universality by showing that all gates of a known universal gate set have a bounded depth bounded fidelity decomposition in the new gate set.


\subsubsection{Designing optimal quantum instruction sets} \label{app:qisa}

YAQQ challenges the canonical gate set used in quantum computation for efficient circuit decomposition.
Though most quantum processors allow generic rotation gates, from the perspective of quantum control, only a discrete gate set can be assigned a finite number of representations.
These can be at either of the 3 levels of (i) quantum assembly, (ii) quantum instruction set architecture~(QISA), or (iii) quantum pulses.
Thus, with YAQQ, we can determine which settings of the rotation angles (of, say, the generic P1 gate) can be hardcoded as a specific quantum instruction.
This is an important problem in the design of the quantum microarchitectures~\cite{fu2019eqasm}.
Similar to how the \{H1, T1\} 1 qubit universality is proven by the SKT, and we understand their importance for entanglement generation and magic-state, the novel gate sets might lead us to discover new properties~\cite{slowik2023calculable} of quantum information processing (e.g., super golden gates~\cite{parzanchevski2018super} and special perfect entanglers~\cite{rezakhani2004characterization}) and aid in the understanding~\cite{cruz2021towards} of quantum resources~\cite{sarkar2022qksa}.
Specifically, it would be crucial to explore the universal distribution~\cite{bach2023visualizing} of quantum states from these gate sets and how the results compare with circuit~\cite{wyeth2023circuit} and algorithmic information theory~\cite{abrahao2022emergence,svozil1995quantum}.



\begin{figure}[!ht]
         \centering
         \includegraphics[clip, trim=0cm 0cm 0cm 0cm,width=\textwidth]{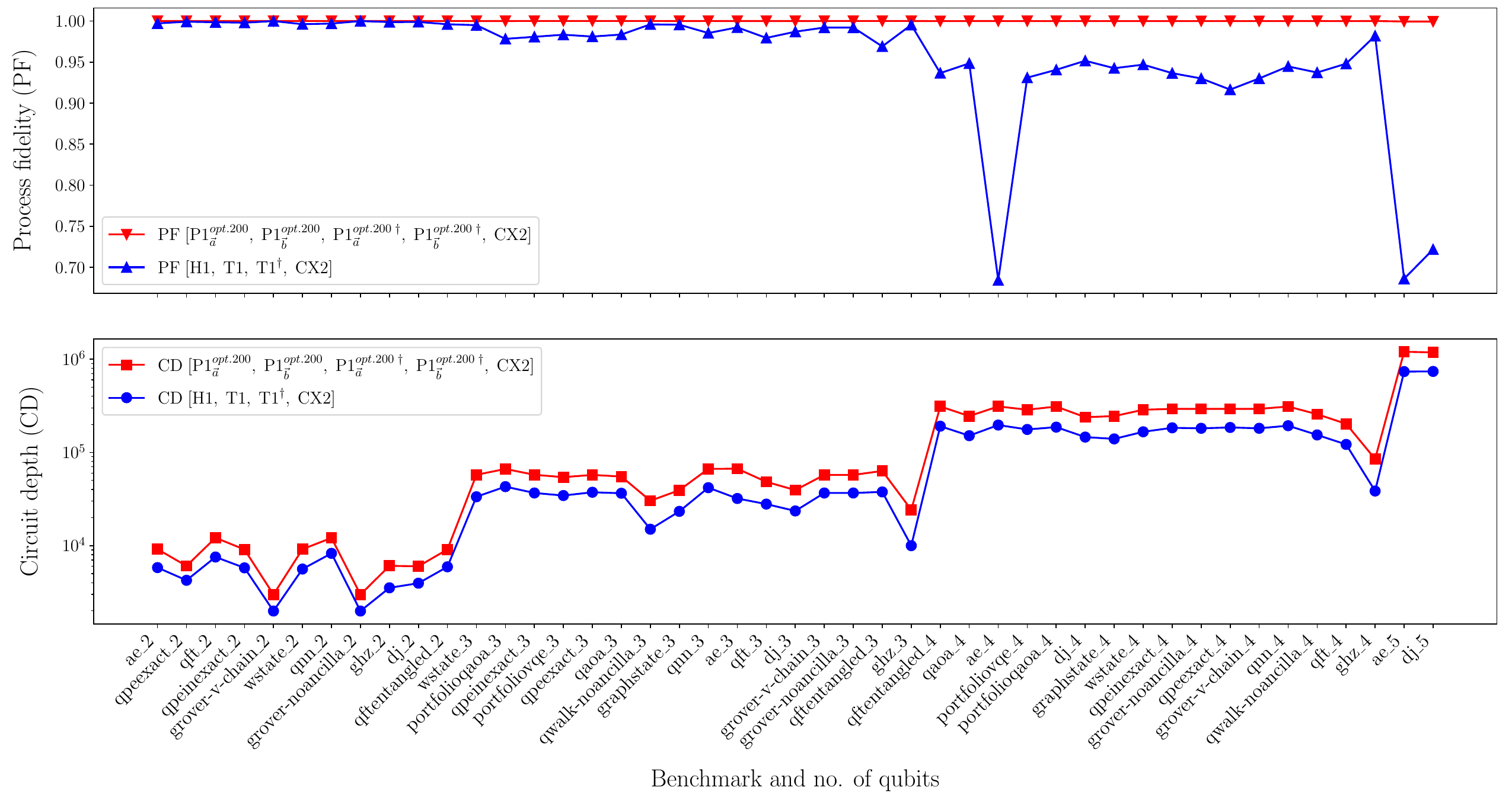}
         \caption{Process fidelity (PF) and circuit depth (CD) of \{$\text{P1}^{opt.200}_{\vec{a}}$, $\text{P1}^{opt.200}_{\vec{b}}$, CX2\} vs. \{H1, T1, CX2\} for 45 unitaries of real-world quantum algorithm from MQT Bench.}
         \label{fig:mqt_yaqq}
\end{figure}

On the other hand, using an over-specified gate set, we can discover new concepts composed of the original gate set; for example, H followed by CX is often used to entangle qubits and thus can be a useful gate that reduces the circuit depth by 1 for every usage.
Such exploration has been done before in \cite{sarra2023discovering}.
In an ongoing research~\cite{QART}, we use YAQQ to optimize the QISA encoding, inspired by similar work in classical assembly targeted for low-power embedded systems.
This set of gates (or quantum instructions) optimizes the quantum instruction bandwidth and energy budget between the quantum processor and quantum compiler.
As a proof-of-concept for this work, we used the optimal gate sets based on 200 Haar random unitaries to decompose $45$ unitaries of well-known quantum algorithms from the MQT Bench~\cite{quetschlich2023mqtbench}.
The comparative process fidelity and circuit depth of using the \{$\text{P1}^{opt.200}_{\vec{a}}$, $\text{P1}^{opt.200}_{\vec{b}}$\} gate set with respect to the standard \{H1, T1\}, (with both gate sets augments with the CX2 gate) for decomposition is shown in Figure~\ref{fig:mqt_yaqq}.
We see similar trends as the 1- and 2-qubit random unitary experiments discussed earlier in Figures~\ref{fig:benchmark_200_points} and \ref{fig:novel_10_points_2q}.
This confirms that the dataset results from random dataset experiments of YAQQ can be utilized in known usage of quantum processors.
Moreover, it confirms that the \{$\text{P1}^{opt.200}_{\vec{a}}$, $\text{P1}^{opt.200}_{\vec{b}}$, CX2\} gate set performs surprisingly better than the standard \{H1, T1, CX2\} gate set in decomposition fidelity, while maintaining a slightly higher circuit depth.
Thus, this gate set can be further studied for QEC and pulse-level specification.

Quantum architecture search~(QAS) is used widely in variational quantum algorithms for ansatz circuit design~\cite{du2022quantum,kundu2024enhancing}.
These ansatz can be defined as custom parametric gates for comparison. 
Similarly, quantum circuit learning~\cite{mitarai2018quantum} directly infers a quantum circuit for a required use case.
By incorporating similar concept discovery and decompilation techniques, YAQQ can be used to aid the design of quantum algorithms and ansatz~\cite{sadhu2024quantum} in the future.


\subsubsection{Compiling to specific quantum processors} \label{app:pulse}

NISQ-focused quantum compilers often support arbitrary gate sets instead of the canonical gates of an error-corrected fault-tolerant quantum computation~(FTQC).
For example, \cite{moro2021quantum,chen2022efficient,wang2022quantumnas} uses techniques from deep reinforcement learning and evolutionary algorithms to tackle the complexity of generic compilation.
In future work, machine learning approaches can be incorporated as an alternate strategy for decomposition within YAQQ.

Various research has targeted the optimization of specific quantum gates ubiquitous in quantum algorithms for specific quantum hardware backends.
For example, within nitrogen-vacancy centers in diamond-based and superconducting Josephson junctions, \cite{goerz2015optimizing} optimizes the implementation of the perfect entangler.
Similar exploration has been done for optimizing the decomposition of the 3-qubit Toffoli gate~\cite{bowman2023hardware} for specific hardware.
The Berkeley Quantum Synthesis Toolkit (BQSKit)~\cite{osti_1785933,kalloor2024quantum} provides various target hardware and runtime optimized gate synthesis and compilation tools, like QFAST, QSearch, LEAP, and QFactor.
Similar to YAQQ, this was used to compare the native gate sets of available QPUs\cite{davis2022gate} based on a set of unitaries.
However, it does not allow the inverse problem of inferring a quantum gate set from the target unitaries.
The BQSKit also enables the bring-your-own-gateset (BYOG) functionality on Infleqtion's Superstaq~\cite{campbell2023superstaq}.
With other hardware-optimized strategies (like dynamic decoupling, routing, and pulse schedules), Superstaq provides a holistic quantum compilation~\cite{lin2022let} solution.
Like Superstaq, YAQQ is being integrated within the OpenQL programming language via programming abstractions for generic unitary decomposition~\cite{krol2022efficient}. 

Our current implementation determines the fidelity based on the process distance between the target and the generated unitary.
Measurement operators can be included as part of the gate set for optimization to account for tomographic completeness.
This type of compilation has been shown to be possible (using the causaloid framework) and implemented using machine learning~\cite{hardy2020quantum}.
YAQQ can be extended to perform a novelty search on a tomographically complete set of maps.
Some ongoing projects are extending YAQQ with these capabilities while maintaining a copyleft AGPL-v3 license.

\begin{figure}[!ht]
         \centering
         \includegraphics[width=\textwidth]{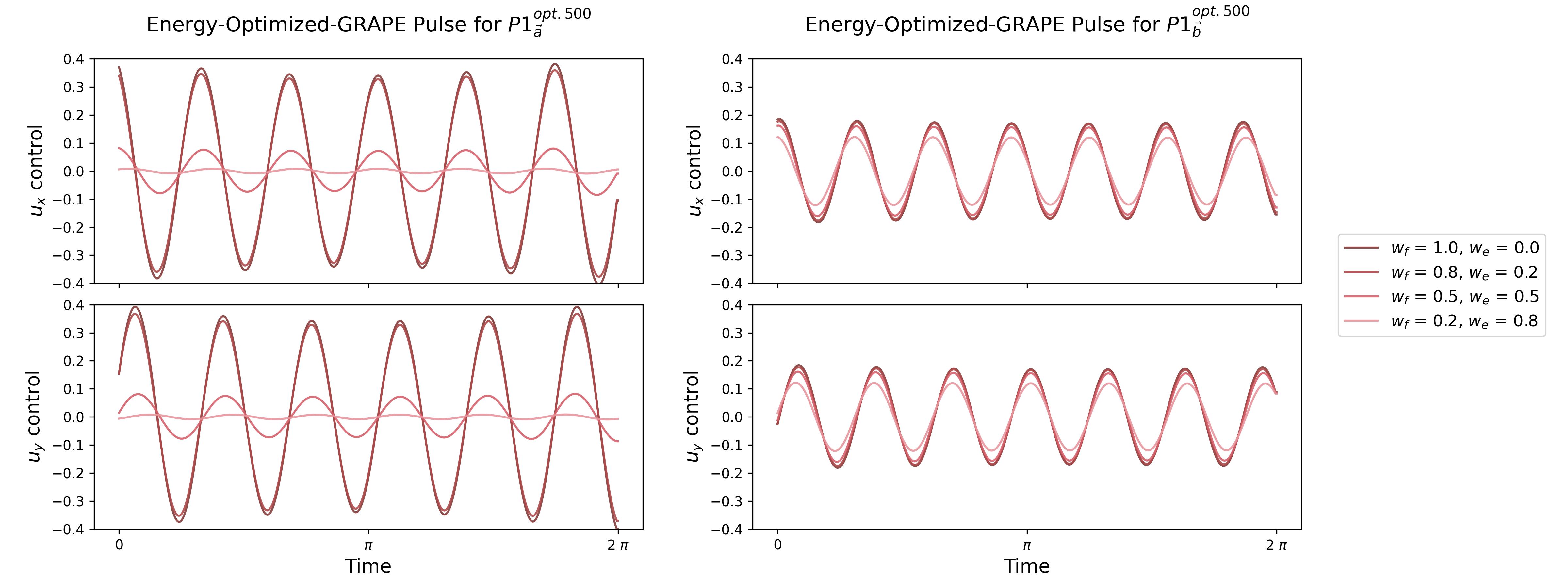}
         \caption{Optimized pulse for \{$\text{P1}^{opt.500}_{\vec{a}}$, $\text{P1}^{opt.500}_{\vec{b}}$\} with trade off between energy and fidelity.}
         \label{fig:eogrape_500}
\end{figure}

The optimal gates suggested by YAQQ can be further optimized at the pulse level for target backends and control configuration.
In ongoing research, we used the optimal gates suggested by YAQQ within an energy-optimized gradient ascent pulse engineering~(EO-GRAPE) framework.
The results for a drift Hamiltonian of $\dfrac{\hbar * \omega}{2} * \sigma_z$ with $\sigma_x$ and $\sigma_y$ control Hamiltonians generate the pulses shown in Figure~\ref{fig:eogrape_500}.
For this experiment, we set $\dfrac{\hbar * \omega}{2} = \pi$.
The fidelity and energy of the generated pulse can be tuned in EO-GRAPE.
The best fidelity achieved is $F(P1^{opt.500}_{\vec{a}}) = 0.9989$, $F(P1^{opt.500}_{\vec{b}}) = 0.99985$

In the future, YAQQ can be integrated with EO-GRAPE in a closed-loop, such that the pulse-level fidelity of a specific quantum hardware configuration can be optimized in the DSE of the gate set.

\section{Conclusion} \label{section:conclusion}
 

This article introduces the tool YAQQ (Yet Another Quantum Quantizer).
It enables the comparison of quantum gate sets by decomposing a set of specified quantum unitaries.
YAQQ can be used as a generic quantum compiler, as well as for benchmarking quantum processors based on their native gate sets.
More importantly, this comparison capability allows YAQQ to perform design space explorations of quantum gate sets.
A novel gate set can be specified as a collection of fixed, parametric, or random gates, which YAQQ can thereafter optimize with respect to another specified gate set.


The formal rationale underlying YAQQ is grounded in algorithmic information theory and quantum universality.
It thus allows for comparing universal and sub-universal computational models with similar theoretical expressibility but different practical reachability.
This subjective evaluation is based on the set of target algorithms or the required quantum transformations for a specific use case and between two gate sets.

In this article, after detailing the various steps involved in this process, we presented experiments demonstrating some insights about quantum gate sets and circuit compilation.
We found 2 sets of gate sets, based on 200 and 500 random unitaries, respectively, which perform better than conventional gate sets on the tuned multi-modal cost function comprising of process fidelity of the approximate decomposition, the quantum circuit depth, and a novelty score.
These performance results were shown to generalize to larger unitaries and real-world quantum algorithms, proving the efficacy of the YAQQ DSE method.
Thereafter, a few exemplary use cases of YAQQ are presented.
This includes comparing resources for different quantum error correction codes, optimizing quantum instruction sets for a set of quantum algorithms, and optimizing quantum control at the pulse level.
These applications will be more thoroughly studied in our future work.
Furthermore, the time-consuming DSE within YAQQ can be accelerated on GPUs or by using deep reinforcement learning-based models.
In light of the surveyed functionalities, we expect the open-sourced YAQQ package to become an indispensable tool for quantum computer architects both in the resource-constrained NISQ era and the discrete-gate-set-constrained FTQC era.

%% file: appendix.tex
\section{Review of decomposition algorithms} \label{a1}

Here we review additional details of the decomposition methods presented in Section~\ref{sec:decomp_tech}.

\subsection{Solovay-Kitaev decomposition}

The chosen set of fundamental gates and the group generated by the set, that is, the SK-basis, must fulfill the following conditions generalized for an $m$-qubit system:
\begin{enumerate}[nolistsep,noitemsep]
    \item All the gates in the set belong to the group of special unitary matrices $SU(m)$ and have a determinant $1$.
    \item The set of gates is closed under inversion, implying that for every gate in the set, its hermitian conjugate must also belong to the set.
    \item The group generated by the set must densely span the space $SU(m)$. This means that, for every arbitrary unitary operation $U$, there must exist a product sequence of gates from the set that can approximate $U$ with a bounded error $\epsilon$.
\end{enumerate}

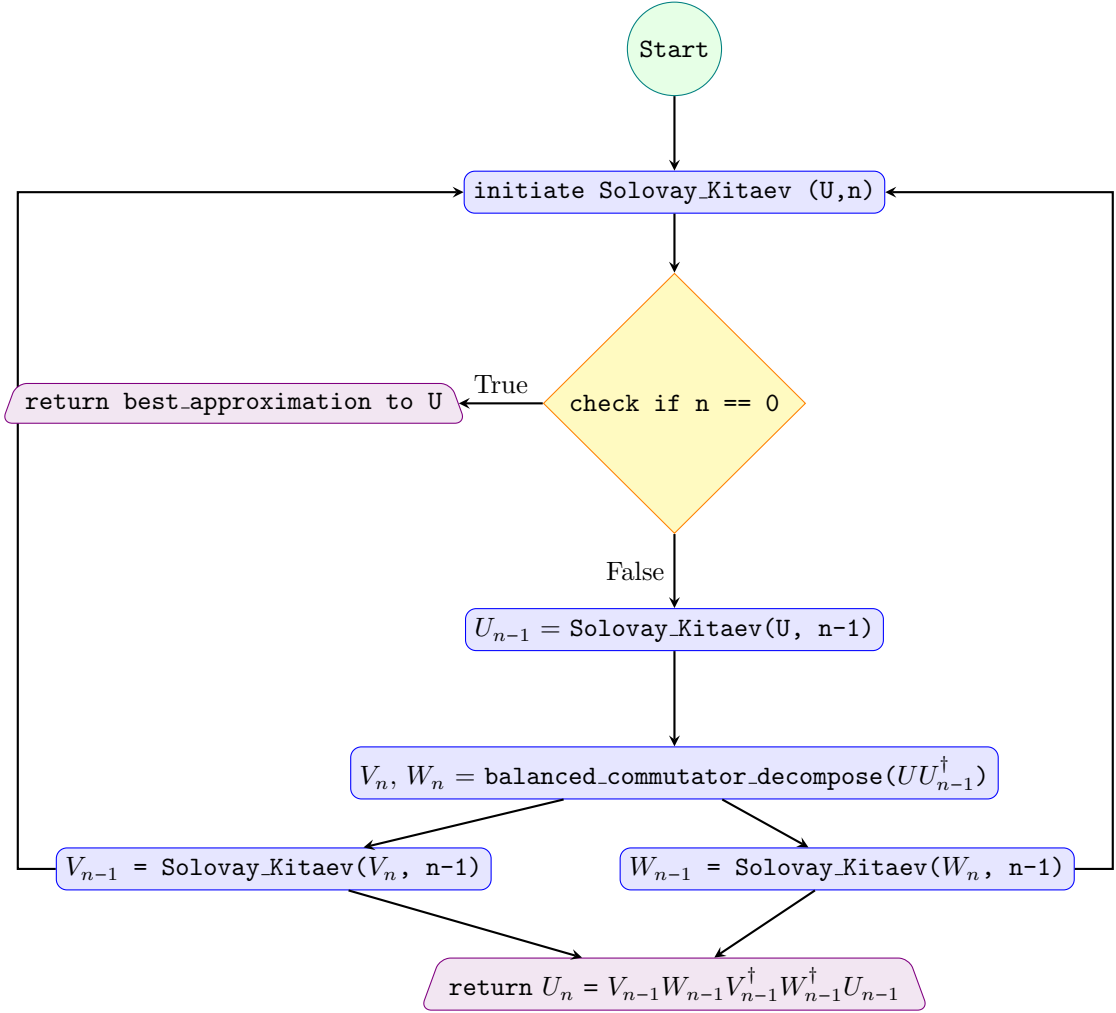
\begin{figure}[!ht]
    \centering
    \input{figs/SK_flowchart}
    \caption{Flowchart of the Solovay-Kitaev Decomposition Algorithm for 1-qubit unitary quantum operator $U$ and recursion depth $n$. Returns the $\epsilon_n$ approximation to the target unitary $U$ computed from call of the function at the $n-1$ degree of recursion, and returns the $\epsilon_0$ approximation in the base case}
    \label{tikz: flowchart}
\end{figure}

The Flowchart~\ref{tikz: flowchart} explains the implementation of SKD based on the pseudocode of the algorithm presented in \cite{dawson2005solovay} and its implementation on \texttt{qiskit}.

The \texttt{balanced\_commutator\_decompose} method performs a balanced group commutator decomposition of the accuracy at level $r$ defined as $\Delta = UU_{r-1}^\dagger = VWV^\dagger W^\dagger$ for matrices $V$ and $W$. $(r-1)$ level approximation accuracies are computed by the call of the function again for matrices $V$ and $W$ and the $r^\text{th}$ level approximate sequence $U_r$ = $V_{r-1}W_{r-1}V_{r-1}^\dagger W_{r-1}^\dagger U_{r-1}$, consisting of all 5 terms computed form the $(r-1)^\text{th}$ level is returned.

\begin{figure}[!ht]
    \centering
    \input{figs/SK-tree}
    \caption{Tree of sequences depicting the working of the Solovay-Kitaev algorithm. Recursion levels are indicated on the left.}
    \label{tikz: skt_tree}
\end{figure}
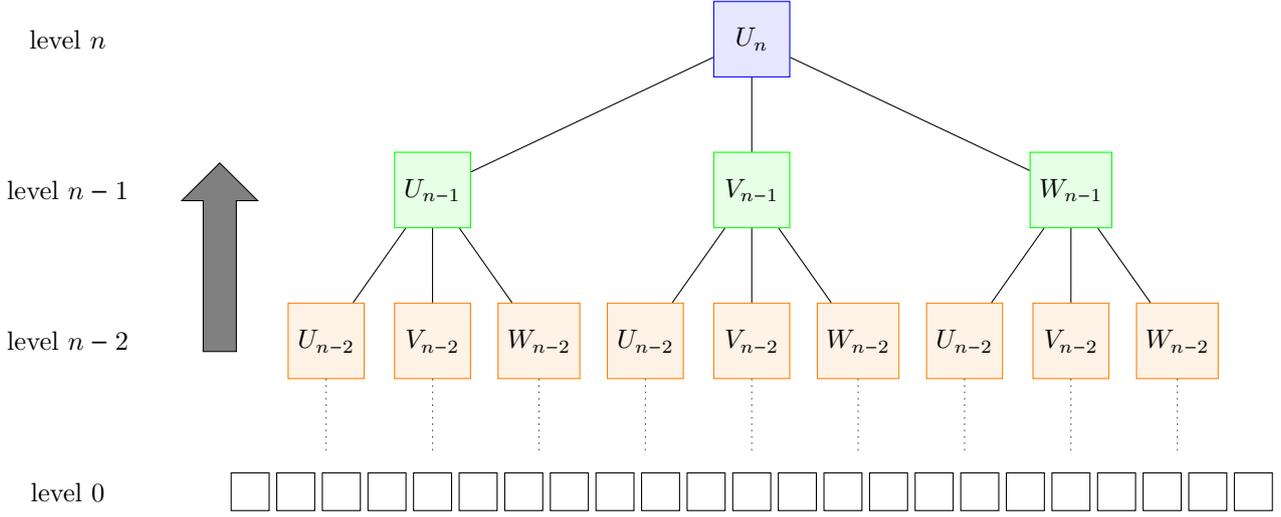

An intuitive tree diagram depicting the working of the algorithm is presented in fig. \ref{tikz: skt_tree}. The bottom-most layer consists of $3^n$ blocks and corresponds to the base case of the algorithm. For each block, the \texttt{best\_approximation} method finds the sequence of gates from the search space that approximates it up to $\epsilon_0$. Each node in the level $r$, $U_r$ is a composite sequence constructed from its three daughter nodes in the $r-1$ level such that $U_r$ = $V_{r-1}W_{r-1}V_{r-1}^\dagger W_{r-1}^\dagger U_{r-1}$. 

\subsection{Canonical decomposition}


A general $2$-qubit gate corresponds to a $4\times 4$ unitary matrix, with $16$ free parameters, i.e., $15$ parameters and an irrelevant global phase. 
Fortunately, any $2$-qubit gate can be decomposed into a $3$-parameter canonical gate, plus $4$ local $3$-parameter (rotation angles across 3 axes) $1$-qubit gates.
Decomposition to the canonical gate is also known as the magic-, Kraus-Cirac-,
or KAK-decomposition.
This decomposing to the canonical representation involves a similarity transformation to the magic basis.
$$V = M\;U\;M^\dagger$$
where $M$ is the magic gate,
$$M = \dfrac{1}{\sqrt{2}}\begin{bmatrix}
    1 & i & 0 & 0 \\
    0 & 0 & i & 1 \\
    0 & 0 & i & -1 \\
    1 & -i & 0 & 0
\end{bmatrix}$$
$M$ diagonalizes the canonical gate as:
$$\text{NL2}_{\vec{t}} = \text{Can}(t_x,t_y,t_z) = M\;D\;M^\dagger$$
and, $D = \text{diag}(e^{i \frac{1}{2}(+t_x-t_y+t_z)},e^{i \frac{1}{2}(-t_x+t_y+t_z)},e^{i \frac{1}{2}(+t_x+t_y-t_z)},e^{i \dfrac{1}{2}(-t_x-t_y-t_z)})$.

If $U$ is a special orthogonal matrix (i.e, Real, $U^T = U$, and $\det U = 1$), then in the magic basis $U$ is the Kronecker product of two $1$-qubit gates, i.e., 
$$V = M\;U\;M^\dagger = A \otimes B \hspace{2em} \text{if } U \in SO(4)$$
We assume that the phase has already been extracted and that $U$ is, therefore, special unitary.
Thus, we can write, $U = (K_3 \otimes K_4)\; \text{Can}(t_x,t_y,t_z)\; (K_1 \otimes K_2)$, and thus:
$$V = M\;U\;M^\dagger =  M(K_3 \otimes K_4)M^\dagger M \text{Can}(t_x,t_y,t_z) M^\dagger M (K_1 \otimes K_2)M^\dagger = Q_2\;D\;Q_1$$

A transpose of $V$ inverts the orthogonal matrices but leaves the complex diagonal matrix unchanged. 
Thus, $V^T V = Q_1^T\;D\;Q_2^T Q_2\;D\;Q_1 =  Q_1^T\;D\;Q_1$ is a similarity transform of the diagonal matrix squared. 
An eigen-decomposition of this yields the square eigenvalues of $D$, and $Q_1$ as the matrix of eigenvectors. 
The canonical gate coordinates can then be extracted from the eigenvalues, and the magic basis transform can be undone to recover the local gates. 
These Kronecker products of local gates can then be decomposed into separate $1$-qubit gates using the Kronecker decomposition and further into elementary gates using a 1-qubit decomposition.

The canonical gate can be further decomposed into CNOT gates or other sets of $2$-qubit gates.
For example, any general canonical gate can be built from 3 CNOT gates as,
\begin{center}
   \includegraphics[height=2cm]{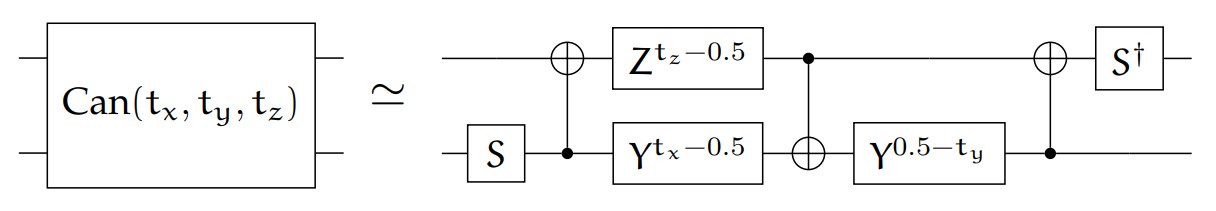} 
\end{center}

Gates on the bottom surface of the Weyl chamber (special orthogonal local equivalency class) require only 2 CNOT gates,
\begin{center}
   \includegraphics[height=2cm]{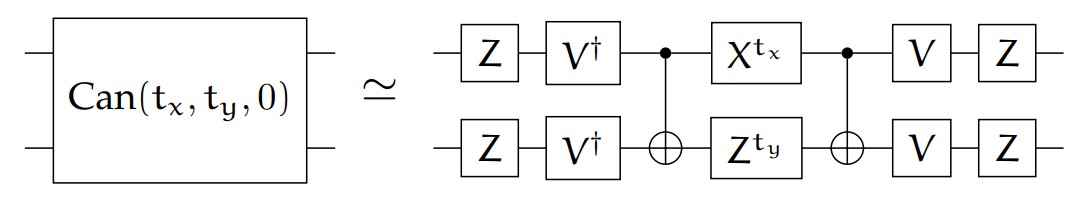} 
\end{center}

While gates locally equivalent to CNOT require only one CNOT gate
\begin{center}
   \includegraphics[height=2cm]{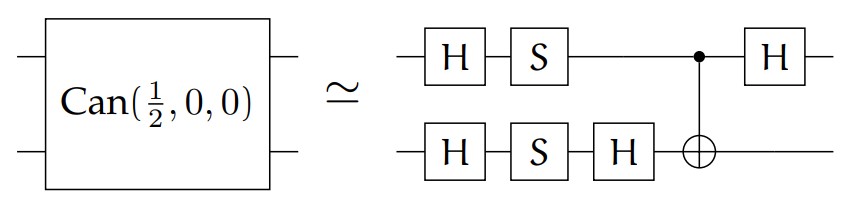} 
\end{center}
and those locally equivalent to the identity require none.

\subsection{Quantum Shannon decomposition}

QSD recursively implements cosine-sine decomposition~(CSD) and other techniques, such as eigenvalue decomposition and Euler decomposition, that divide the target matrix $U$ into smaller blocks. 
Eventually, the unitary is expressed as a sequence of single-qubit rotation gates and CNOTs.

According to CSD, $U = LMR^\dagger$ where $L$ and $R$ are block-diagonal matrices representing uniformly controlled gates and the middle matrix $M$ representing a controlled $R_y$ rotation on the MSB. 
\begin{equation}
    \label{csd}
    U = 
    \begin{bmatrix}
        \begin{array}{c|c}
        U_{00} & U_{01} \\ \hline
        U_{10} & U_{11}
        \end{array}
    \end{bmatrix}
    = 
    \begin{bmatrix}
        \begin{array}{c|c}
        L_1 & 0 \\ \hline
        0 & L_2
        \end{array}
    \end{bmatrix}
    \begin{bmatrix}
        \begin{array}{c|c}
        C & -S \\ \hline
        S & C
        \end{array}
    \end{bmatrix}
    \begin{bmatrix}
        \begin{array}{c|c}
        R_1 & 0 \\ \hline
        0 & R_2
        \end{array}
    \end{bmatrix}^\dagger
\end{equation}
$L_1$, $L_2$, $R_1$ and $R_2$ are unitary matrices of size $2^{n-1}$. $C$ and $S$ are diagonal matrices such that $C^2+S^2 = I$, thereby justifying the name of the decomposition technique. The matrices $L$ and $R$ are termed as quantum multiplexors, and they enact $L_1$ ($R_1$) or $L_2$ ($R_2$) conditioned on the state of the MSB. The middle matrix resembles the $R_y$ rotation matrix that is targeted on the MSB and controlled by the states of the lower-order qubits.


\begin{figure}[!ht]
    \centering
    \captionsetup{justification=centering}
    \includegraphics[width=8cm]{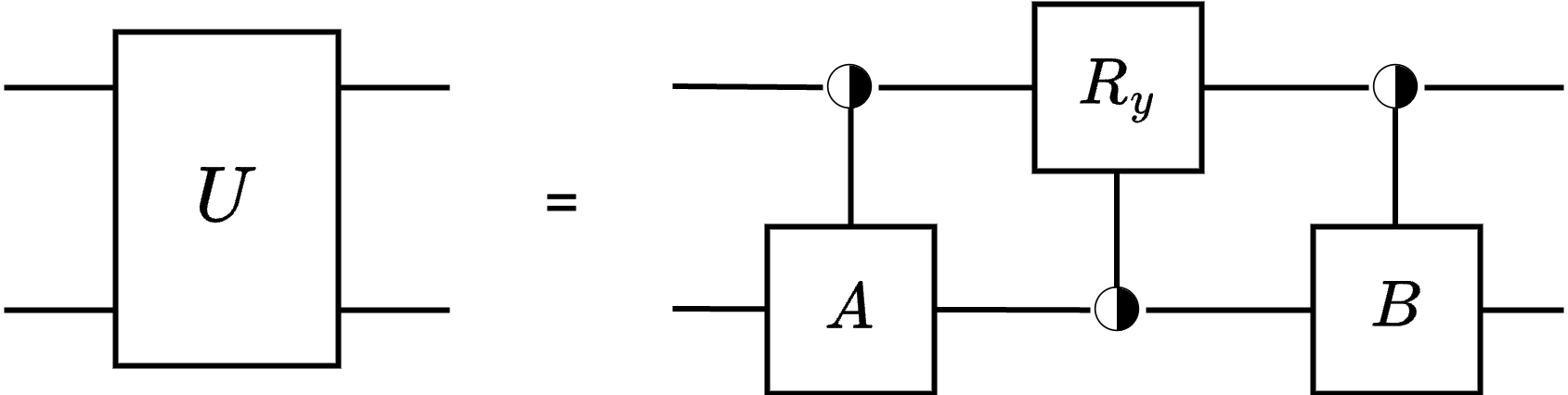}
    \caption{The Cosine-Sine decomposition acting on $n$-qubit gate $U$. The slash represents a bundle of wires, and the box control symbol indicates multiple control wires}
    \label{cosine-sine decomposition}
\end{figure}

The left and right gates undergo a demultiplexing routine that performs an eigenvalue decomposition of the matrices.
\begin{equation}
    \label{demux}
    \begin{bmatrix}
        A_1 & 0 \\ 
        0 & A_2
    \end{bmatrix}
    = 
    \begin{bmatrix}
        P & 0 \\ 
        0 & P
    \end{bmatrix}
    \begin{bmatrix}
        \Lambda & 0 \\ 
        0 & \Lambda^\dagger
    \end{bmatrix}
    \begin{bmatrix}
        Q & 0 \\ 
        0 & Q
    \end{bmatrix}
\end{equation}
$P$ and $Q$ are unitary matrices, and $\Lambda$ is a unitary diagonal matrix. The left and right matrices are quantum gates operating on the lower-order qubits and are independent of the MSB. The middle matrix corresponds to a $R_z$ operation on the MSB controlled by the lower qubits.

\begin{figure}[!ht]
    \centering
    \captionsetup{justification=centering}
    \includegraphics[width=5cm]{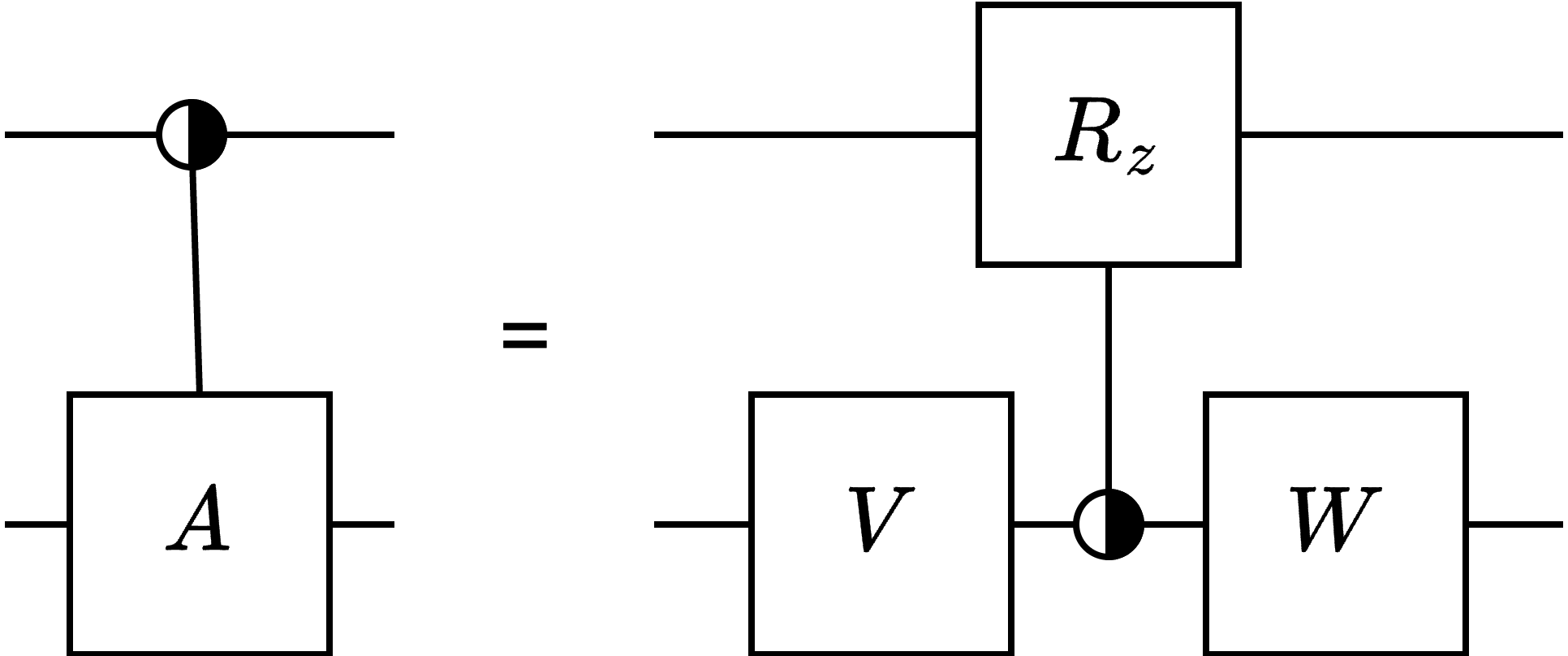}
    \caption{Demultiplexing of a multiplexor}
    \label{qsd_demultiplex}
\end{figure}

This process of CSD, followed by subsequent demultiplexing operation, is performed recursively until the algorithm reaches the base case. 
At the base level, the operator sequence consists of only single qubit gates. 
At this point, any single-qubit unitary operation is changed into a rotation gate following Euler decomposition. 
The implementation of the algorithm can have two options, $a1$ and $a2$. 
In $a1$, the multiplexed $R_y$ operation in \ref{cosine-sine decomposition} is implemented using $CZ$ gates, while in $a2$, the recursion is stopped at the level of 2-qubit operations, and the resulting circuit is decomposed into CNOT gates and single-qubit rotation gates. 

\begin{figure}[!ht]
    \centering
    \captionsetup{justification=centering}
    \includegraphics[width=12cm]{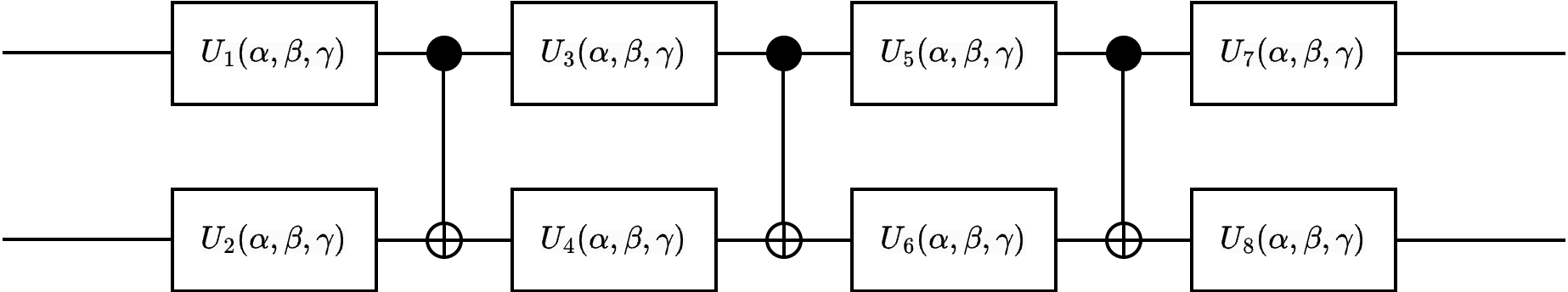}
    \caption{Final decomposed form of a 2-qubit arbitrary unitary operator expressed as a sequence of 1-q rotations and CNOT gates.}
    \label{2q_QSD}
\end{figure}

In this project, the \texttt{qiskit} implementation of the QSD algorithm is employed with the $a2$ option.
It returns a decomposed quantum circuit consisting of single qubit unitary rotations and CNOT gates as depicted in Figure~\ref{2q_QSD} for a 2-qubit system.

\section{Definitions of standard built-in gates} \label{a2}

The novel gate set ansatz can be specified based on the available gate options, as explained in Section~\ref{s2p5}.
The exact definitions of these gates are given in Table~\ref{gates}.

\begin{table}[!ht]
\centering
\caption{Elements for composing gate sets}
\label{gates}
\begin{tabular}{ |p{1.4cm}|p{12cm}|p{2.2cm}| } 
    \hline
    Gate Identifier & Description & Search Method \\ 
    \hline
    R1 & 1-qubit Haar random unitary gate & RND  \\ 
    P1 & 1-qubit parametric gate (IBM U3) & RND/OPT(3)  \\ 
    T1 & 1-qubit T gate 
    $:= \bigg[\begin{smallmatrix}
        1 & 0 \\
        0 & e^{i\pi/4}
    \end{smallmatrix}\bigg]$
    & FXD  \\ 
    TD1 & 1-qubit T-dagger gate
    $:= \bigg[\begin{smallmatrix}
        1 & 0 \\
        0 & e^{-i\pi/4}
    \end{smallmatrix}\bigg]$ & FXD  \\ 
    S1 & 1-qubit S gate 
    $:= \bigg[\begin{smallmatrix}
        1 & 0 \\
        0 & e^{i\pi/2}
    \end{smallmatrix}\bigg]$
    & FXD  \\ 
    Z1 & 1-qubit Z gate 
    $:= \bigg[\begin{smallmatrix}
        1 & 0 \\
        0 & e^{i\pi}
    \end{smallmatrix}\bigg]
    = \bigg[\begin{smallmatrix}
        1 & 0 \\
        0 & -1
    \end{smallmatrix}\bigg]$
    & FXD  \\ 
    X1 & 1-qubit X gate 
    $:= \bigg[\begin{smallmatrix}
        0 & 1 \\
        1 & 0
    \end{smallmatrix}\bigg]$
    & FXD  \\ 
    H1 & 1-qubit Hadamard gate
    $:= \frac{1}{\sqrt{2}} \bigg[ \begin{smallmatrix}
        1 & 1 \\
        1 & -1
    \end{smallmatrix}\bigg]$ & FXD  \\ 
    F1 & 1-qubit unitary gate definition from file & depends  \\ 
    R2 & 2-qubit Haar random unitary gate & RND  \\ 
    NL2 & 2-qubit non-local unitary gate & RND/OPT(3)  \\ 
    CX2 & 2-qubit CNOT gate
    $:= \Bigg[ \begin{smallmatrix}
        1 & 0 & 0 & 0 \\
        0 & 1 & 0 & 0 \\
        0 & 0 & 0 & 1 \\
        0 & 0 & 1 & 0 
    \end{smallmatrix} \Bigg]$
    & FXD  \\ 
    B2 & 2-qubit Berkeley gate
    $:= \Bigg[ \begin{smallmatrix}
        \cos(\pi/8) & 0 & 0 & i\sin(\pi/8) \\
        0 & \cos(3\pi/8) & i\sin(3\pi/8) & 0 \\
        0 & i\sin(\pi/8) & \cos(\pi/8) & 0 \\
        i\sin(\pi/8) & 0 & 0 & \cos(\pi/8) 
    \end{smallmatrix} \Bigg]$ & FXD  \\ 
    SPE2 & 2-qubit special perfect entangler $:= \text{NL2}(0.5,t_y,0)$ & RND/OPT(1)  \\ 
    F2 & 2-qubit unitary gate definition from file & depends  \\ 
    \hline
\end{tabular}
 
\end{table}

%% file: figs/SK_flowchart.tex
\pgfdeclarelayer{background}
\pgfsetlayers{background,main}
\tikzstyle{startstop} = [circle, minimum width=0.5cm, minimum height=0.5cm,text centered, draw=teal, fill=green!10]
\tikzstyle{io} = [trapezium, rounded corners, trapezium left angle=70, trapezium right angle=70, minimum width=1cm, minimum height=0.5cm, text centered, draw=violet, fill=violet!10]
\tikzstyle{process} = [rectangle, rounded corners, minimum width=1cm, minimum height=0.5cm, text centered, draw=blue, fill=blue!10]
\tikzstyle{decision} = [diamond, minimum width=1cm, minimum height=0.5cm, text centered, draw=orange, fill=yellow!30]
\tikzstyle{arrow} = [thick,->,>=stealth]

\begin{tikzpicture}[node distance=1.8cm]
    \node (start) [startstop] {\texttt{Start}};
    \node (initiate) [process, below of=start, yshift=-0.1cm] {\texttt{initiate Solovay\_Kitaev (U,n)}};
    \node (check) [decision, below of=initiate, yshift=-1cm] {\texttt{check if n == 0}};
    \node (skcall) [process, below of=check, yshift=-1.2cm] {$U_{n-1}$ = \texttt{Solovay\_Kitaev(U, n-1)}};
    \node (decompose) [process, below of=skcall, yshift=-0.1cm] {$V_n$, $W_n$ = \texttt{balanced\_commutator\_decompose(}$UU^{\dagger}_{n-1}$\texttt{)}};
    \node (skv) [process, below left of=decompose, xshift=-4.0cm] {$V_{n-1}$\texttt{ = Solovay\_Kitaev(}$V_n$\texttt{, n-1)}};
    \node (skw) [process, below right of=decompose, xshift=1.0cm] {$W_{n-1}$\texttt{ = Solovay\_Kitaev(}$W_n$\texttt{, n-1)}};
    \node (return) [io, below of=decompose, yshift=-1cm] {\texttt{return} $U_{n} = V_{n-1}W_{n-1}V_{n-1}^{\dagger}W^{\dagger}_{n-1}U_{n-1}$};
    \node (bestapprox) [io, left of=check, xshift=-4cm] {\texttt{return best\_approximation to U}};
    
    \draw [arrow] (start) -- (initiate);
    \draw [arrow] (initiate) -- (check);
    \draw [arrow] (check) -- node[anchor=east] {False} (skcall);
    \draw [arrow] (check) -- node[anchor=south] {True} (bestapprox);
    \draw [arrow] (skcall) -- (decompose);
    \draw [arrow] (decompose) -- (skv);
    \draw [arrow] (decompose) -- (skw);
    \draw [arrow] (skw.east) -- ++(0.5,0) |- (initiate.east);
    \begin{pgfonlayer}{background}
    \draw [arrow] (skv.west) -- ++(-0.5,0) |- (initiate.west);
    \end{pgfonlayer}
    \draw [arrow] (skv) -- (return);
    \draw [arrow] (skw) -- (return);
    
    \end{tikzpicture}

%% file: figs/SK-tree.tex
\tikzstyle{lvl1} = [rectangle, minimum width=1cm, minimum height=1cm, text centered, draw=blue, fill=blue!10]
\tikzstyle{lvl2} = [rectangle, minimum width=1cm, minimum height=1cm, text centered, draw=green, fill=green!10]
\tikzstyle{lvl3} = [rectangle, minimum width=1cm, minimum height=1cm, text centered, draw=orange, fill=orange!10]
\tikzstyle{lvl0} = [rectangle, minimum width=0.5cm, minimum height=0.5cm, text centered, draw=black]
\tikzstyle{upwardarrow} = [
  single arrow,                 
  draw,                         
  fill=gray,                   
  minimum height=2.5cm,         
  minimum width=1cm,            
  single arrow head extend=.1cm,
  shape border rotate=90,       
  inner sep=2pt                 
]

\begin{tikzpicture}[node distance=2cm]
    \node (start) [lvl1] {$U_n$};
    \node (txt_start) [left of=start, xshift=-7cm]{level $n$};
    \node (un-1) [lvl2, below of=start, xshift=-4.2cm] {$U_{n-1}$};
    \node (vn-1) [lvl2, below of=start] {$V_{n-1}$};
    \node (wn-1) [lvl2, below of=start, xshift=4.2cm] {$W_{n-1}$};
    \node (txt_start) [left of=vn-1, xshift=-7cm]{level $n-1$};
    \node (uun-2) [lvl3, below of=un-1, xshift=-1.4cm] {$U_{n-2}$};
    \node (vun-2) [lvl3, below of=un-1] {$V_{n-2}$};
    \node (wun-2) [lvl3, below of=un-1, xshift=1.4cm] {$W_{n-2}$};
    \node (uvn-2) [lvl3, below of=vn-1, xshift=-1.4cm] {$U_{n-2}$};
    \node (vvn-2) [lvl3, below of=vn-1] {$V_{n-2}$};
    \node (txt_start) [left of=vvn-2, xshift=-7cm]{level $n-2$};
    \node (wvn-2) [lvl3, below of=vn-1, xshift=1.4cm] {$W_{n-2}$};
    \node (uwn-2) [lvl3, below of=wn-1, xshift=-1.4cm] {$U_{n-2}$};
    \node (vwn-2) [lvl3, below of=wn-1] {$V_{n-2}$};
    \node (wwn-2) [lvl3, below of=wn-1, xshift=1.4cm] {$W_{n-2}$};
    \node (base) [lvl0, below of=uun-2, xshift=-1cm]{};
    \node (base) [lvl0, below of=uun-2, xshift=-0.4cm]{};
    \node (base) [lvl0, below of=uun-2, xshift=0.2cm]{};
    \node (base) [lvl0, below of=uun-2, xshift=0.8cm]{};
    \node (base) [lvl0, below of=uun-2, xshift=1.4cm]{};
    \node (base) [lvl0, below of=uun-2, xshift=2.0cm]{};
    \node (base) [lvl0, below of=uun-2, xshift=2.6cm]{};
    \node (base) [lvl0, below of=uun-2, xshift=3.2cm]{};
    \node (base) [lvl0, below of=uun-2, xshift=3.8cm]{};
    \node (base) [lvl0, below of=uun-2, xshift=4.4cm]{};
    \node (base) [lvl0, below of=uun-2, xshift=5.0cm]{};
    \node (baselol) [lvl0, below of=uun-2, xshift=5.6cm]{};
    \node (txt_start) [left of=baselol, xshift=-7cm]{level $0$};
    \node (base) [lvl0, below of=uun-2, xshift=6.2cm]{};
    \node (base) [lvl0, below of=uun-2, xshift=6.8cm]{};
    \node (base) [lvl0, below of=uun-2, xshift=7.4cm]{};
    \node (base) [lvl0, below of=uun-2, xshift=8.0cm]{};
    \node (base) [lvl0, below of=uun-2, xshift=8.6cm]{};
    \node (base) [lvl0, below of=uun-2, xshift=9.2cm]{};
    \node (base) [lvl0, below of=uun-2, xshift=9.8cm]{};
    \node (base) [lvl0, below of=uun-2, xshift=10.4cm]{};
    \node (base) [lvl0, below of=uun-2, xshift=11.0cm]{};
    \node (base) [lvl0, below of=uun-2, xshift=11.6cm]{};
    \node (base) [lvl0, below of=uun-2, xshift=12.2cm]{};
    \node (arrow) [upwardarrow, left of=vvn-2, xshift=-5cm, yshift=1cm]{};

    \draw[-] (start) -- (un-1);
    \draw[-] (start) -- (vn-1);
    \draw[-] (start) -- (wn-1);
    \draw[-] (un-1) -- (uun-2);
    \draw[-] (un-1) -- (vun-2);
    \draw[-] (un-1) -- (wun-2);
    \draw[-] (vn-1) -- (uvn-2);
    \draw[-] (vn-1) -- (vvn-2);
    \draw[-] (vn-1) -- (wvn-2);
    \draw[-] (wn-1) -- (uwn-2);
    \draw[-] (wn-1) -- (vwn-2);
    \draw[-] (wn-1) -- (wwn-2);
    \draw[dotted] (uun-2.south) -- ++ (0,-1);
    \draw[dotted] (vun-2.south) -- ++ (0,-1);
    \draw[dotted] (wun-2.south) -- ++ (0,-1);
    \draw[dotted] (uvn-2.south) -- ++ (0,-1);
    \draw[dotted] (vvn-2.south) -- ++ (0,-1);
    \draw[dotted] (wvn-2.south) -- ++ (0,-1);
    \draw[dotted] (uwn-2.south) -- ++ (0,-1);
    \draw[dotted] (vwn-2.south) -- ++ (0,-1);
    \draw[dotted] (wwn-2.south) -- ++ (0,-1);
    
    \end{tikzpicture}